\documentclass[%
 reprint,
 amsmath,amssymb,
 aps,
]{revtex4-1}

\usepackage{multirow,booktabs}
\usepackage[table]{xcolor}
\usepackage{graphicx}% Include figure files
\usepackage{dcolumn}% Align table columns on decimal point
\usepackage{bm}% bold math
\usepackage[breaklinks]{hyperref} %make citations and cross-references clickable; allow links to be split between different lines
\usepackage[all]{hypcap}%click to figure or table instead of figure or table caption (needs hyperref)
\usepackage{relsize}
\def\cesrta{{C{\smaller[2]ESR}TA}}

\begin{document}

%\preprint{APS/123-QED}

\title{Investigation into electron cloud effects in the International Linear Collider positron damping ring}
%\author{J.~A.~Crittenden, J.~Conway, G.~F.~Dugan, M.~A.~Palmer}
%\altaffiliation{Present address: Fermi National Accelerator Laboratory, P.O.Box 500, Batavia, IL 60510-5011}
%\author{D.~L.~Rubin, J.~Shanks, K.~G.~Sonnad}
\author{J.~A.~Crittenden{$^\dagger$}, J.~Conway, G.~F.~Dugan, 
M.~A.~Palmer{$^{\ddagger}$}, D.~L.~Rubin, J.~Shanks, and K.~G.~Sonnad}
%\email{crittenden@cornell.edu}
\affiliation{CLASSE, Cornell University, Ithaca, NY 14853, USA}
\author{L.~Boon and K.~Harkay} 
\affiliation{ANL, Argonne, IL 60439, USA}
\author{T.~Ishibashi}
\affiliation{KEK, 1-1 Oho, Tsukuba, Ibaraki 305-0801, Japan}
\author{M.~A.~Furman}
\affiliation{LBNL, Berkeley, CA 94720, USA}
\author{S.~Guiducci}
\affiliation{INFN Laboratori Nazionali di Frascati, P.O. Box 13, I-00044, Frascati (Roma), Italy}
\author{M.~T.~F.~Pivi}
\altaffiliation{\vskip -8.9mm \hspace*{-6.3mm}
{$^{\ddagger}$} \hspace*{0mm} \begin{minipage}[t]{\columnwidth}{Present address: Fermi National Accelerator Laboratory,\\ 
                       P.O.Box 500, Batavia, IL 60510-5011}\end{minipage}\\*[.2mm]
Present address: IMS Nanofabrication, Vienna, Austria\\*[.2mm]
{\hspace*{-3.3mm} $^{\dagger}$ crittenden@cornell.edu}
}
\affiliation{SLAC, Menlo Park, CA 94025, USA}
\author{L.~Wang}
\affiliation{SLAC, Menlo Park, CA 94025, USA}

\date{\today}% It is always \today, today,

\begin{abstract}
We report modeling results for electron cloud buildup and instability in the 
International Linear Collider positron damping ring. 
Updated optics, wiggler magnets, and vacuum chamber designs
have recently been developed for the 5~GeV, 3.2-km racetrack layout. 
An analysis of the synchrotron
radiation profile around the ring has been performed, including the effects of 
diffuse and specular
photon scattering on the
interior surfaces of the vacuum chamber. The results provide input to the cloud buildup 
simulations for the various magnetic field regions of the ring. 
The modeled cloud densities thus obtained are used in the instability threshold 
calculations. We conclude that the mitigation techniques employed in this model 
will suffice to allow operation of the damping ring at the 
design operational specifications.
\end{abstract}

\pacs{Valid PACS appear here}% PACS, the Physics and Astronomy
\maketitle

\section{Introduction}
The discoveries  at the Large Hadron Collider~\cite{ref:higgsatlas,ref:higgscms} 
have re-intensified interest 
in the proposed International Linear Collider (ILC)~\cite{ref:ilctdrex}. 
Operation of the ILC depends 
critically on the reliable performance of the electron and positron damping rings (DRs) 
which will serve as injectors. Electron cloud (EC) buildup has been 
shown to limit the performance of
storage rings at KEK-B~\cite{ref:eckekb} and PEP-II~\cite{ref:ecpepii}, the operating parameters
of which are comparable to those of the ILC DRs.
For the past several years, we have been developing and validating modeling codes for the 
purpose of designing the ILC DRs. This paper presents the results of those 
efforts. We present the beam optics design, the vacuum chamber designs including 
recommended cloud buildup mitigation techniques, cloud buildup simulations and 
modeling estimates of the effects on beam dynamics, 
deriving conclusions on the feasibility of building and 
operating the positron DR to specification.

\section{Design of the Positron Damping Ring Lattice}
\begin{figure}[tbh]
        \includegraphics[width=0.99\columnwidth]{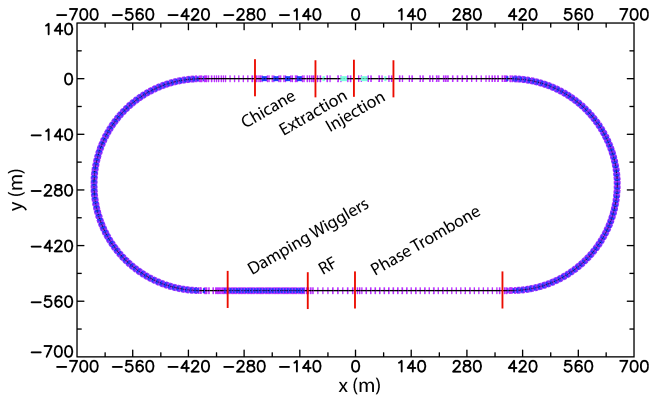}
        \caption{Layout of DTC04 lattice}
        \label{fig:dtc04_floor}
    \end{figure}

The lattice design used for the EC buildup and instability simulations is the so-called 
DTC03 lattice, with arc cells 
designed by Rubin et al.~\cite{ref:IWFLC12:Shanks} and straights based on the work of 
Korostelev and Wolski~\cite{ref:IPAC10:WEPE096}. 
The lattice has since undergone minor revisions to improve matching between the straights and arc sections, and has 
iterated to DTC04~\cite{ref:ilctdracc}. 
The differences between DTC03 and DTC04 are insignificant for the purposes of the studies described here. 
The racetrack layout for the 3238-m circumference ring
is shown in Fig.~\ref{fig:dtc04_floor}.
The  100-m-long
RF straights can accommodate as many as 16~single-cell cavities and the 226-m wiggler 
straight
up to 60~superferric wiggler magnets~\cite{ref:ipac12jacwig}. 

\begin{figure}[tbh]
        \includegraphics[width=0.99\columnwidth]{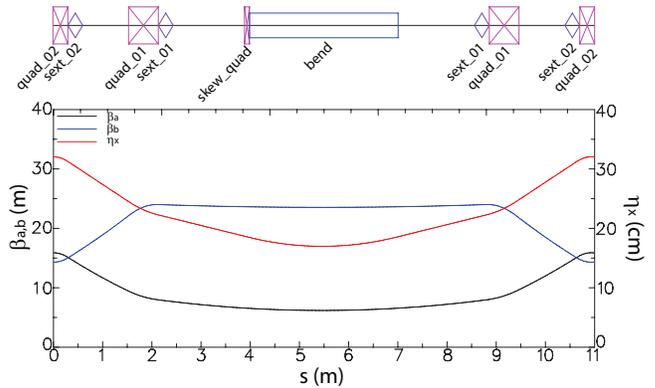}
        \caption{Horizontal and vertical beta functions $\beta_{\rm a}$ and $\beta_{\rm b}$, 
                 and the horizontal
                 dispersion function $\eta_{\rm x}$ in the DTC04 arc cell}
        \label{fig:dtc04_arccell}
    \end{figure}

\begin{figure*}[tbh]
        \includegraphics[width=0.99\textwidth]{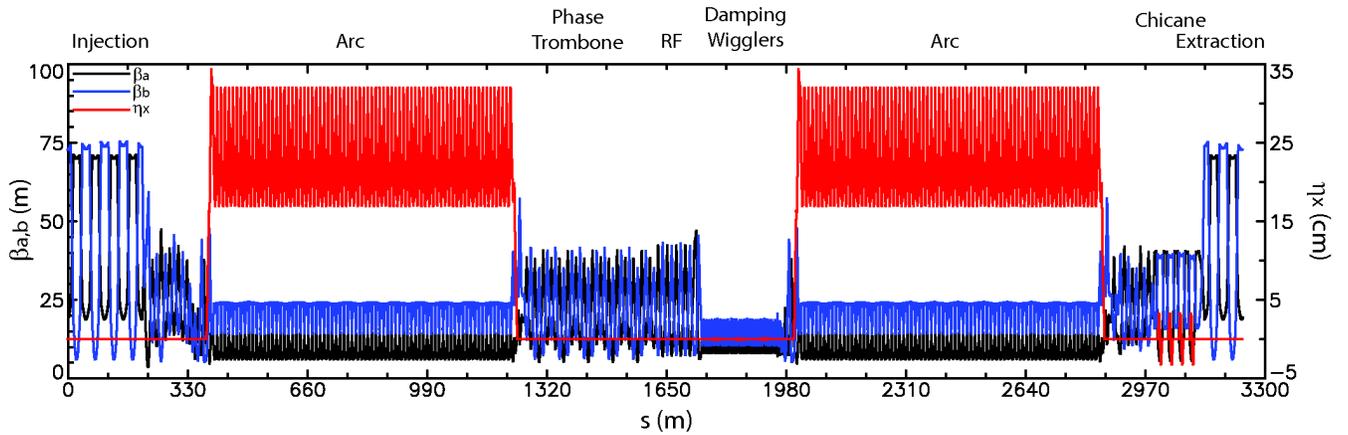}
        \caption{Horizontal and vertical beta functions $\beta_{\rm a}$ and $\beta_{\rm b}$, 
                 and the horizontal
                 dispersion function $\eta_{\rm x}$ for the entire DTC04 lattice}
        \label{fig:dtc04_twiss}
    \end{figure*}

The operational parameters of the damping ring are given in Tab.~\ref{tab:dtc04_summary}. 
        \begin{table}[htb]
        \begin{centering}
            \caption {Summary of the DTC04 lattice parameters}
            \label{tab:dtc04_summary}
            \begin{tabular}{lcc}
                \toprule[1pt]
                \addlinespace[2pt]
                \textbf{Parameter} & \textbf{Value} & \textbf{Units}\\
                \midrule[0.5pt]
                Circumference & 3238 & m\\
                Energy      & 5.0 & GeV \\
                Betatron Tunes ($Q_{\rm x}$, $Q_{\rm y}$)& (48.850, 26.787) & \\
%                Tune Advance per Arc  & (16.418, 6.074) & \\
                Chromaticity ($\xi_{\rm x},\xi_{\rm y}$) & (1.000, 0.302)   & \\
%                Chromaticity per Arc  & (9.074, 10.896) & \\
                Train Repetition Rate & 5 & Hz \\
                Minimum Bunch Spacing   & $6.15$ & ns \\
                Bunch Population      & $2 \times 10^{10}$ &  \\
                Extracted $\epsilon_{\rm x}^{\rm geometric}$ & 0.6 & nm\\
                Extracted $\epsilon_{\rm y}^{\rm geometric}$  & $<2$ & pm\\
                Extracted Bunch Length  & 6  & mm \\
                Extracted $\sigma_E/E$ & 0.11 & \% \\
                Damping Time      & 24  & ms \\
                Wiggler $B^{\rm max}$ & 1.5 & T \\
                \bottomrule[1pt]
            \end{tabular}
        \end{centering}
        \end{table}
The baseline design 
(26-ms damping time and 5-Hz operation)
requires 8~cavities with total accelerating voltage of 14~MV and 54~2.1-m-long 
wiggler magnets with 1.51-T peak field.
In order to run in the proposed 10-Hz mode, the wigglers operate at 2.16~T to cut the 
radiation damping time in half,
and the accelerating voltage is increased to 22.4~MV with 12~cavities to preserve 
the 6-mm bunch length.
The 339-m phase trombone in the wiggler straight
consists of five six-quadrupole cells 
and has a tuning
range of $\pm0.5$ betatron wavelengths. The opposite straight includes injection and 
extraction lines,
and the 117-m-long chicane for fine adjustment of the revolution period. The range of the 
chicane is
${\pm}4.5$~mm with negligible contribution to the horizontal emittance. The arc cell, 
shown in Fig.~\ref{fig:dtc04_arccell}, is a 
simple variation of
a TME-style cell with a single 3-m bend, three quadrupoles (one focusing and two 
defocusing), four sextupoles, a skew quadrupole and two beam position monitors.
Figure~\ref{fig:dtc04_twiss} shows the beta functions and horizontal dispersion function for
the entire DTC04 lattice.
There are 75 cells in each arc. A complete list of components is given in Tab.~\ref{tab:mag_count}. 
    \begin{table}[htb]
        \begin{centering}
        \caption {Summary of elements in the DTC04 lattice}
        \label{tab:mag_count}
        \begin{tabular}{lr}
            \toprule[1pt]
            \addlinespace[2pt]
            \textbf{Class} & \textbf{Count} \\
            \midrule[0.5pt]
            Beam Position Monitor  & 511 \\
            Dipole          & 164 \\ % includes chicane and B01
            Horizontal Steering & 150 \\
            Vertical Steering  & 150 \\
            Combined H+V Steering  & 263 \\
            Quadrupole      & 813 \\
            Skew Quadrupole & 160 \\
            Sextupole       & 600 \\
            Damping Wigglers & 54    \\
            \bottomrule[1pt]
        \end{tabular}
        \end{centering}
    \end{table}
The dynamic aperture including magnet multipole errors 
and misalignments, and
wiggler nonlinearities, is large enough to accept 
an injected positron phase space with
normalized horizontal and vertical emittances $A_{\rm x}$ and $A_{\rm y}$ 
such that \mbox{$A_{\rm x} + A_{\rm y} < 0.07$~mrad} and energy spread
$\Delta E/E \le 0.075\%$~\cite{ref:ipac12jsh}.

\section{Vacuum Chamber Design}
\begin{figure}[tb]
   \centering
   \includegraphics*[width=0.99\columnwidth]{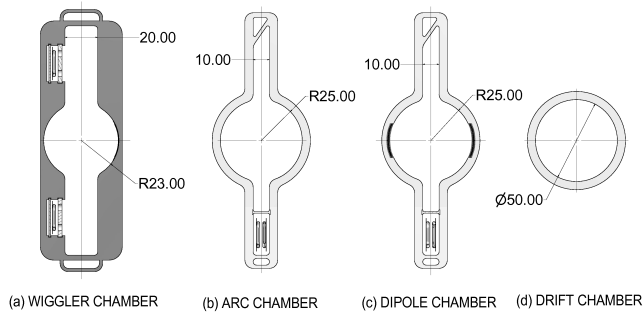}
   \caption{ \label{fig:vcprofiles}
            Vacuum chamber profiles for the a)~wiggler magnets, b)~arc sections, c)~dipole magnets,
and d)~the fieldfree regions of the damping ring. Note the positions of the NEG strips, the grooves
in the dipole vacuum chambers and the angled rear walls of the antechambers.
}
\end{figure}

\begin{figure}[tb]
   \centering
   \includegraphics*[width=0.99\columnwidth]{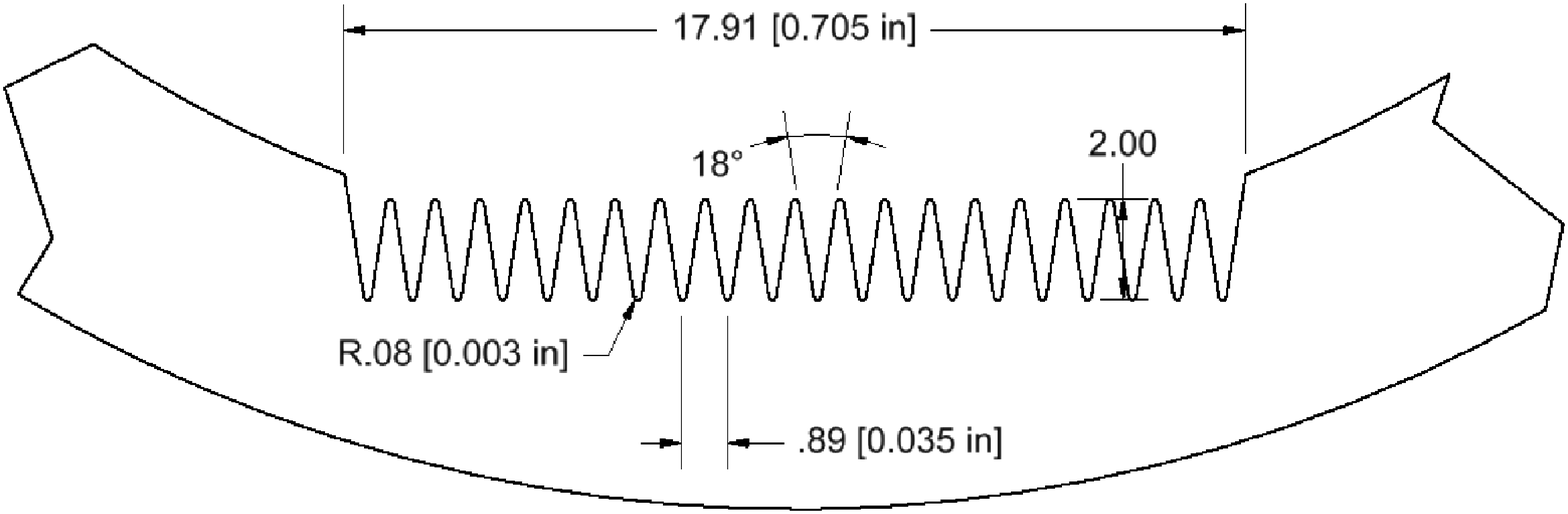}
   \caption{ \label{fig:Dipole_Grooves}
            Schematic cross section of bottom wall showing the 
            dimensions of the grooves used to suppress EC buildup
            in the dipole magnets
}
\end{figure}

\begin{figure}[tb]
   \centering
   \includegraphics*[width=0.99\columnwidth]{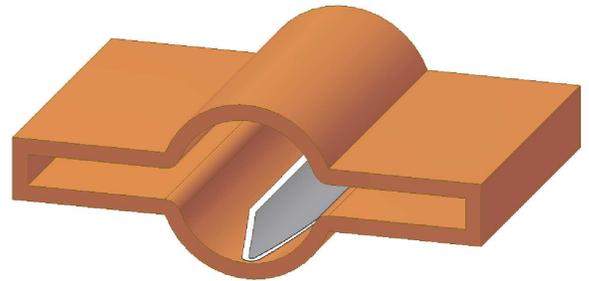}
   \caption{ \label{fig:Wiggler_Electrode}
            Wiggler magnet vacuum chamber showing the clearing electrode
}
\end{figure}

\begin{figure}[tb]
   \centering
   \includegraphics*[width=0.99\columnwidth]{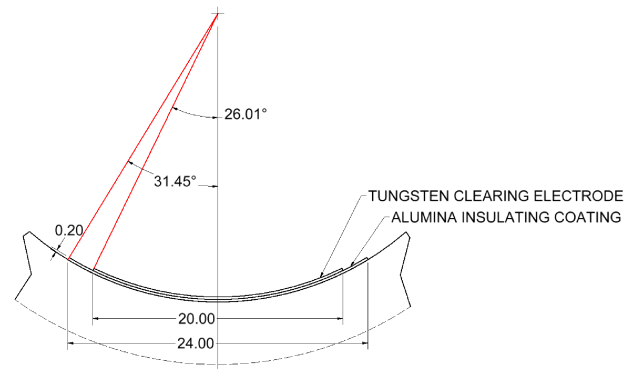}
   \caption{ \label{fig:Clearing_Electrode}
            Dimensions of the clearing electrode designed for cloud suppression in the 
            wiggler magnets
}
\end{figure}

The conceptual design of the vacuum chambers incorporates mitigation techniques
in each of the various magnetic field 
environments to suppress the local buildup of the EC.
The mitigation methods were selected based on the results of an intense
research effort conducted as part of the ILC Technical Design program~\cite{ref:ipac11_pivi}.  The vacuum system
conceptual design is described in Ref.~\cite{ref:ipac12jvc}.  
The vacuum chamber profiles chosen for the wiggler, arc, dipole and fieldfree regions
of the ring are shown in Fig.~\ref{fig:vcprofiles}.
In the arc regions of the ring, the 50-mm aperture vacuum
chambers employ a TiN coating to suppress secondary electron yield (SEY)  and
dual antechambers to reduce the number of photoelectrons which can seed the cloud.
The rear walls of the antechambers are angled in order to suppress photon backscattering 
into the beam region.
In the dipoles, the EC is further suppressed by the use of longitudinal grooves
on the top and bottom surfaces, as shown in Figs.~\ref{fig:vcprofiles}~c) and~\ref{fig:Dipole_Grooves}. 
In the wiggler region, a 46-mm aperture chamber 
utilizes clearing
electrodes (see Figs.~\ref{fig:Wiggler_Electrode} and~\ref{fig:Clearing_Electrode}) to suppress growth of the 
cloud and dual antechambers 
along with custom
photon stops to suppress the generation of photoelectrons.  
Drift regions throughout
the ring will employ solenoid windings to further reduce the EC density in the vicinity of the beam.

\section{Photon Transport Model}
The distribution of synchrotron radiation striking the walls of the vacuum chamber
can be used to predict the sources of the photoelectrons which seed the EC.
This distribution has been computed for the ILC DR lattice using
a newly developed photon-tracking simulation code, Synrad3D~\cite{ref:ECLOUD10:PST08}. 
This code computes the synchrotron
radiation photons per positron generated by a beam circulating in the magnetic lattice,
and simulates the propagation  in three dimensions, of the photons as they scatter off,
or are absorbed by, the vacuum chamber. The design vacuum chamber geometry, including
details such as antechambers and photon stops, is used in the calculation.
Both specular and diffuse photon scattering are included in the simulation. For the scattering
calculation, the surface material is approximated as aluminum with a thin carbon coating,
and the surface roughness parameters are typical of a technical vacuum chamber,
namely rms roughness 0.1~$\mu$m and autocorrelation length 5~$\mu$m.

\begin{figure}[tb]
   \centering
   \includegraphics*[width=0.9\columnwidth]{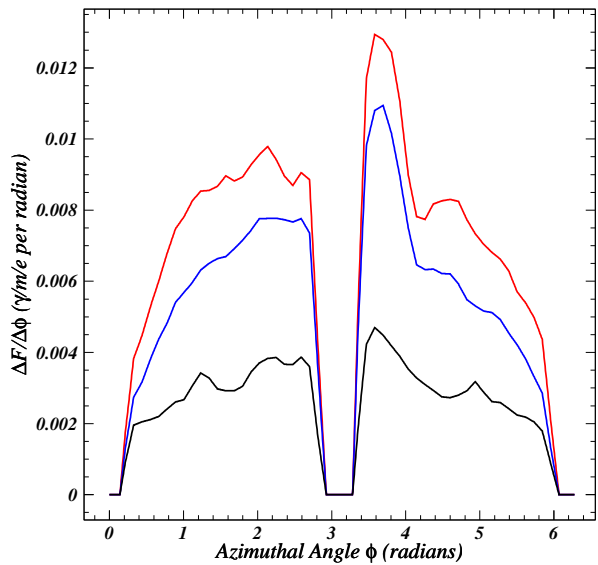}
   \caption{ \label{fig:synrad3d}
Absorbed photon rate per radian $\frac{{\Delta}F}{\strut \Delta{\phi}}(\phi)$
versus azimuthal angle $\phi$ in Arc 1 of the DR, averaged over the regions corresponding to
three types of magnetic environment: 1)~quadrupole fields (red), 2)~fieldfree regions (blue) and
3)~dipole fields (black). 
The azimuthal angle $\phi$ is defined to be zero where the vacuum chamber
intersects the bend plane on the outside of the ring. The angle $\pi /2$ corresponds to the
top of the vacuum chamber. The low photon rates at zero and $\pi$ radians are due to the absorption
in the antechambers.
}
\end{figure}

Figure~\ref{fig:synrad3d}
shows the photon intensity
distributions for magnetic elements in one of the arcs of the DR.
The low photon rates at zero and $\pi$ radians are due to the antechambers.
The top-bottom asymmetry is due to the angle in
the antechamber back walls, which inhibits scattering out of the antechamber.

This photon transport model was also used to calculate the consequences of photon scattering for the vacuum chamber heat load.
In particular, the synchrotron radiation produced by the superconducting wiggler magnets produces intense heating on the 
vacuum chamber walls.  
Photon absorbers are used to shield the central region of the beam chamber in the superconducting 
wigglers~\cite{ref:Malyshev_IPAC10}.  

Each wiggler magnet produces 25.2 kW of synchrotron radiation power. Since the wiggler straight is 200~m long 
most of this power will be absorbed within the wiggler section.  
The photon absorbers were designed to absorb 40~kW of radiation power each~\cite{ref:Zolotarev_IPAC10}. 
The modeled 499~mm-long conical absorbers have diameters varying from 44~mm to 52~mm and are placed
between pairs of damping wiggler magnets. The synchrotron radiation is incident on a 23-mm long tapered
section at the end of the absorber.

\begin{figure}
    \includegraphics[width=0.99\columnwidth]{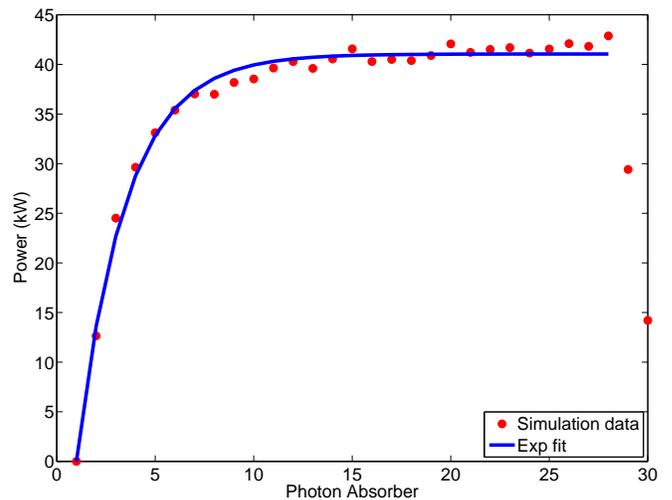}
\caption{Exponential fit to the calculated power dissipated on each photon absorber in the 
damping wiggler 
straight section}
\label{fig:PAHeating_expfit}
\end{figure}

The method described in~\cite{ref:Boon_IPAC11} was used to calculate the total power dissipated in each 
photon absorber.   
We compared the dissipated power for three models of photon scattering: no scattering, 
specular scattering and diffuse scattering.  The first case assumed 
that all photons 
incident on the chamber wall are absorbed, yielding  maximum absorbed power of 40.3 kW.  This result 
agrees with previous 
analytical calculations presented 
in Ref.~\cite{ref:Zolotarev_IPAC10} for the 6.4~km damping ring.  The specular scattering model used a 
reflectivity based 
on a surface roughness of 4~nm rms, 
yielding a maximum power of 42.9 kW.  The model for diffuse scattering assumed the surface roughness 
parameters typical 
of a technical aluminum vacuum chamber, 0.1~$\mu$m rms and 5~$\mu$m autocorrelation length, resulting in 
maximum power of 41.0 kW. 
The calculated absorbed power for this model and an exponential fit are shown in 
Fig.~\ref{fig:PAHeating_expfit}.
We conclude that the absorber design is capable of handling the synchrotron radiation power produced 
by the damping wigglers.

\section{EC Buildup in the Arc Dipoles}
We have employed the code POSINST~\cite{ref:furmanpivi}, to simulate EC buildup in the 
arc dipoles of the ILC DR lattice~\cite{ref:ecloud10maf} under the following assumptions:
1)~the SEY model parameters are those obtained from fits to 
measurements obtained at the
Cornell Electron Storage Ring Test Accelerator ({\cesrta}) project~\cite{ref:icfa50} for a 
TiN surface~\cite{ref:PAC11:TUP230},
2)~the distribution of photons striking the chamber surface at the location of the dipole magnet
has been obtained from Synrad3D calculations including photon scattering, and
3)~the quantum efficiency is assumed to be 0.05, independent of photon energy and incident angle.

The SEY model corresponding to the above-mentioned fits yields a peak SEY value of 0.94 at an 
incident electron energy of 296~eV.
In addition, we have carried out the simulation in which the SEY is set to 0 (meaning that any 
electron hitting
the chamber walls gets absorbed with unit probability) in order to isolate the contribution to the 
EC density $\rm{N}_e$ from photoemission.
The results are summarized in Tab.~\ref{tab:posinst}.
Cloud densities averaged over the full vacuum chamber in the 1-m-long test volume as well as those 
averaged over a 
$20\sigma_x \times 20\sigma_y$ 
elliptical cross-sectional
area centered on the beam axis, where $\sigma_x$ and $\sigma_y$ 
denote the horizontal and vertical rms beam sizes, are shown. The modeling statistical uncertainties 
are less than 30\%.
\begin{table}[htbp]
\centering
\caption{ \label{tab:posinst}
POSINST modeling results for EC densities $\rm{N}_e$ ($10^{11}$~m$^{-3}$) in the dipole regions of the 
ILC DR lattice. 
The first row shows the beam pipe-averaged density 
at the end of a 34-bunch train. The second row shows the peak 
$20\sigma$ density during the train passage. The third row gives the maximum $20\sigma$ density 
just prior to the arrival of any bunch.}
\begin{tabular}{lcc}
\toprule[1pt]
\addlinespace[2pt]
\textbf{SEY} & \textbf{0} & \textbf{0.94} \\
\midrule[0.5pt]
34-bunch density & $0.5$  & $1.2$ \\
Peak $20\sigma$ density & $0.2$  & $0.5$ \\
$20\sigma$ density prior to bunch arrival & $0.2$  & $0.4$ \\
\bottomrule[1pt]
\end{tabular}
\end{table}

The results of the simulation with no secondary electron production provide a lower limit on $\rm{N}_e$,
however, one must bear in mind that this lower limit is directly proportional to the model value for the  quantum efficiency,
here assumed to be~5\%.
For peak SEY=0.94, $\rm{N}_e$ is a factor of 2 or 3 greater than that for SEY=0.
These results for peak SEY=0.94 represent an upper limit, since the effects of the grooves in the 
dipole vacuum chamber design were not accounted for in the simulation.
The $20\sigma$~densities are somewhat smaller than the above-quoted average over the entire vacuum chamber, 
as are the $20\sigma$-densities prior to bunch passage.

The effectiveness of grooves for suppression of EC buildup has been the subject of a number of modeling 
studies~\cite{ref:pac07lw,ref:pac07mv}.
Measurements of the reduction in secondary yield afforded by such grooves have been performed at 
PEP-II~\cite{EPAC08:MOPP064,JAP104:104904}.
More recently, measurements of the reduction of cloud buildup in a grooved aluminum vacuum
chamber relative to that for a smooth chamber surface in the CESR positron storage ring showed 
an improvement by more than a factor of two, corresponding to a decrease in the peak SEY value from 2.0 to 
1.2~\cite{IPAC13:MOPWA072}.

\section{EC Buildup in the Quadrupoles, Sextupoles and Fieldfree Regions}
\begin{figure}[tb]
   \centering
   \includegraphics*[width=0.99\columnwidth]{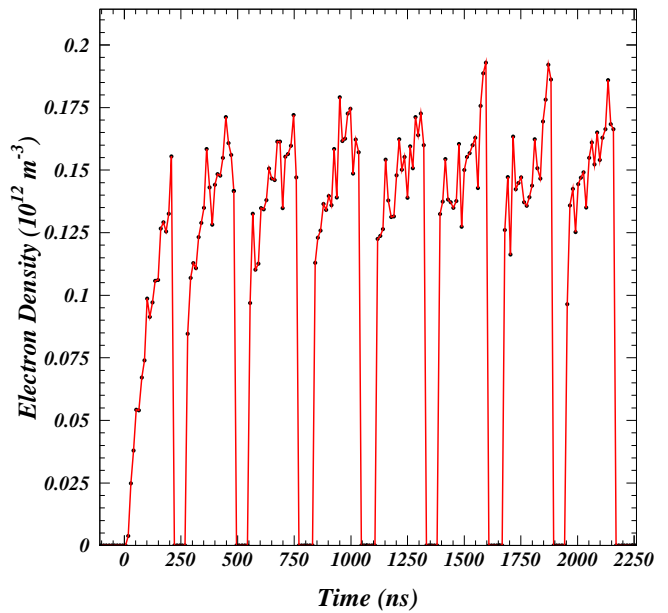}
   \caption{ \label{fig:dens_quad} 
            $20\sigma$ cloud densities on beam axis
            just prior to the passage of each of the 34 bunches in 8 trains in a 
            quadrupole magnet in an arc region of the ILC DR. 
            Cloud trapping effects build up over many train passages, but the maximum central density 
            along a train stabilizes 
            at a value less than $2 \times 10^{11}$~m$^{-3}$ after two trains 
}
\end{figure}

\begin{figure}[tb]
   \centering
   \includegraphics*[width=0.94\columnwidth]{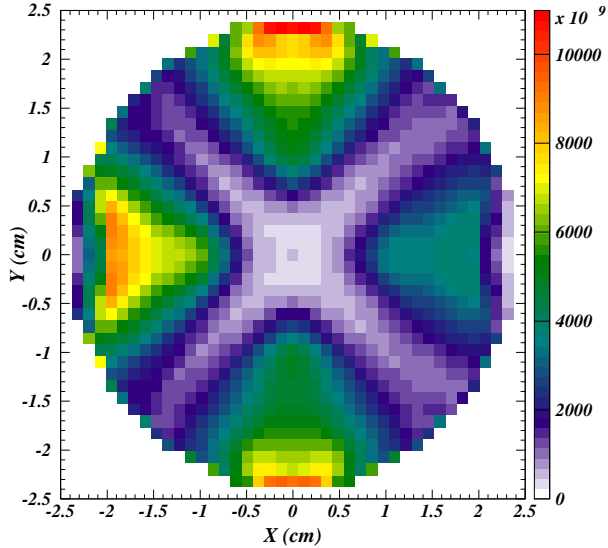}
   \caption{ \label{fig:profile_quad} 
            Simulated EC density profile on a $41 \times 41$ grid 
            in a quadrupole magnet in an arc region of the ILC CR. 
            averaged over the 2.2~$\mu$s simulation time. 
            Cloud electron escape zones are observed along the diagonal regions connecting magnet poles, as 
            are several regions of trapped cloud. 
}
\end{figure}

\begin{figure}[tb]
   \centering
   \includegraphics*[width=0.94\columnwidth]{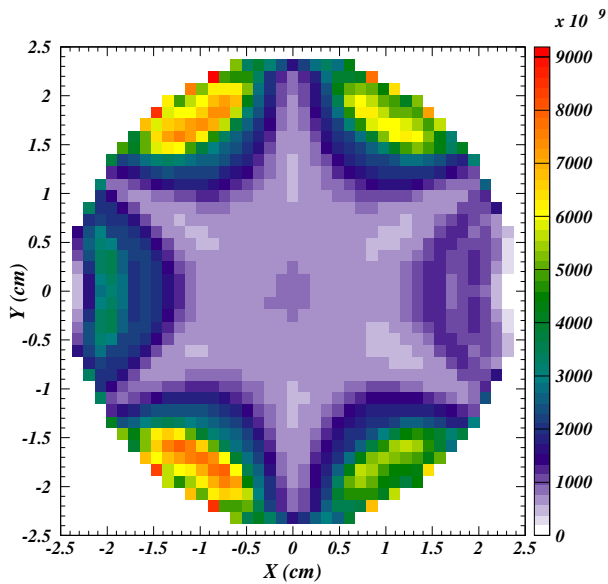}
   \caption{ \label{fig:profile_sext} 
            Simulated EC density profile  averaged over the 2.2~$\mu$s simulation time 
            in an ILC DR sextupole magnet
}
\end{figure}

The EC buildup modeling code ECLOUD~\cite{CERN:SL2002:016AP,PRSTAB4:012801}, served to calculate estimates of the cloud densities in the quadrupoles and sextupoles in the arc and wiggler regions and in the fieldfree regions of the wiggler sections for the ILC DR lattice.
The photon transport modeling code Synrad3D provided photon absorption distributions averaged over
each of these regions. The ECLOUD code was updated to use the POSINST-style photoelectron production and SEY model parameters~\cite{ECLOUD12:Fri1240}. Comparative studies of the ECLOUD and POSINST codes, including validation
with {\cesrta} coherent tune shift measurements in dipole magnets have been presented in 
Refs.~\cite{PAC09:TH5PFP047} and~\cite{IPAC10:TUPD024}. The ECLOUD code was also extended to sextupole magnetic fields for the purposes of this study.
Representative field strengths of 10~T/m (70~T/m$^2$)were used for the quadrupoles (sextupoles). Trapping effects 
were evident
in the beam pipe-averaged cloud densities, which had not yet reached equilibrium after eight train passages, 
but since
the trapping does not occur in the central beam region, the cloud density in the $20\sigma$ beam region
just prior to the passage of each bunch
(shown in Fig.~\ref{fig:dens_quad}) was stable after just a couple of train passages. 
Figure~\ref{fig:profile_quad} shows the cloud density profile averaged over the 2.2~$\mu$s simulation. 
The higher density regions,
including those with long-term trapped cloud, do not populate the beam axis. The $20\sigma$ cloud densities calculated in the
field of a sextupole magnet also reach saturation during the first two trains, 
%field of a sextupole magnet (see Fig.~\ref{fig:dens_sext}) also reach saturation during the first two trains, 
%\begin{figure}[tb]
%   \centering
%   \includegraphics*[width=0.99\columnwidth]{ECLOUD_31393_3.eps}
%   \caption{ \label{fig:dens_sext} 
%            $20\sigma$ cloud density just prior to the passage of each of the 34 bunches in 8 trains in a sextu%pole magnetic field
%}
%\end{figure}
and the density profile  is also depleted on the beam axis, as shown in Fig.~\ref{fig:profile_sext}.

\begin{figure}[tb]
   \centering
   \includegraphics*[width=0.94\columnwidth]{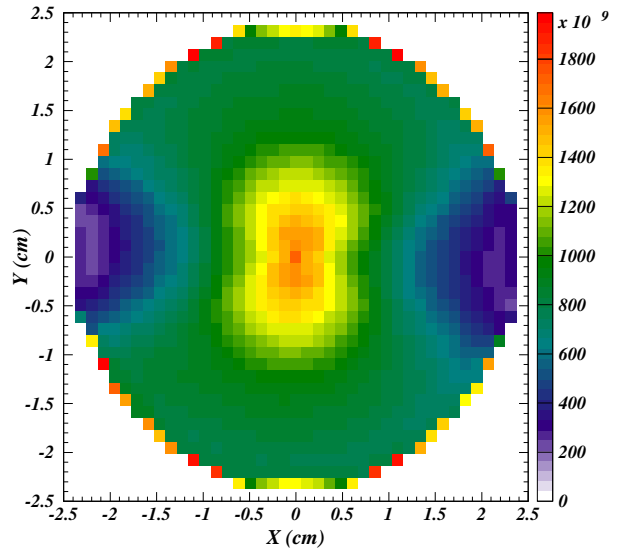}
   \caption{ \label{fig:profile_ff} 
            Simulated EC density profile averaged over the 4.5~$\mu$s simulation time in a 
            fieldfree region of arc region 1 without application of the 
            solenoidal magnetic field mitigation technique
}
\end{figure}

\begin{figure}[tb]
   \centering
   \includegraphics*[width=0.94\columnwidth]{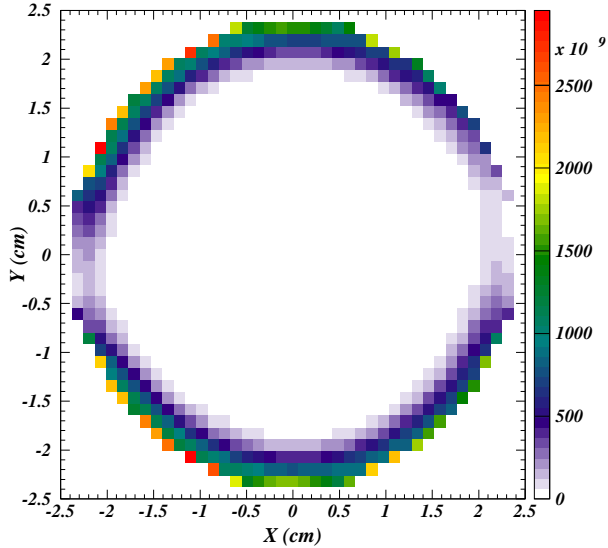}
   \caption{ \label{fig:profile_ffsol} 
            Effect of a 40-G solenoidal magnetic field on the simulated 
            EC density profile in a fieldfree region of the ring 
}
\end{figure}

Table~\ref{tab:ecloud} shows the 20$\sigma$ density estimates prior to each bunch passage obtained 
assuming a peak SEY value of 0.94.
\begin{table}[htbp]
\centering
\caption{ \label{tab:ecloud}
         POSINST and ECLOUD modeling results for the 20$\sigma$ density estimates $\rm{N}_e$ ($10^{11}$~m$^{-3}$) 
         just prior to each bunch passage in the ILC DR lattice design.  
         The total length of each magnetic field environment $L$ is given in meters.}
\begin{tabular}{lcccccccc}
\toprule[1pt]
\addlinespace[2pt]
& \multicolumn{2}{c}\textbf{Fieldfree} & \multicolumn{2}{c}\textbf{Dipole} & \multicolumn{2}{c}\textbf{Quadrupole} & \multicolumn{2}{c}\textbf{Sextupole} \\
& L & $N_e$ & L & $N_e$ & L & $N_e$ & L & $N_e$ \\ 
Arc region 1 & 406 & 2.5 & 229 & 0.4 & 146 & 1.5 & 90 & 1.4 \\
Arc region 2 & 365 & 2.5 & 225 & 0.4 & 143 & 1.7 & 90 & 1.3 \\
Wiggler region & 91 & 40 & 0 & & 18 & 12 & 0 & \\
\bottomrule[1pt]
\end{tabular}
\end{table}
The POSINST results for the arc dipoles are included in this table. The integrated ring lengths for the
magnetic environment types are also shown. The high density values in the quadrupole magnets of the wiggler section
of the ring result from the intense wiggler radiation. The densities are an order of magnitude greater than those in the
arc regions, but the integrated length of those quadrupoles is an order of magnitude smaller.
The simulations for the fieldfree regions were repeated imposing a solenoidal magnetic
field of 40~G, as is foreseen in the mitigation recommendations determined during the ECLOUD10 workshop~\cite{ref:ipac11_pivi}.
Such a field was shown to reduce the cloud buildup in the vicinity of the beam to negligible levels. 
The $20\sigma$ cloud density
at a time immediately prior to the passage of each of the bunches was
found to be $2.5 \times 10^{11}$ with no applied solenoidal field.
%The $20\sigma$ cloud density
%at a time immediately prior to the passage of each of the bunches is shown in 
%Fig.~\ref{fig:dens20_ff} with no solenoidal field.
%\begin{figure}[tb]
%   \centering
%   \includegraphics*[width=0.99\columnwidth]{ECLOUD_31372_3.eps}
%   \caption{ \label{fig:dens20_ff} 
%            $20\sigma$ cloud density just prior to the passage of each of the 34 bunches in 
%            16 trains in a fieldfree region of arc region 1.
%}
%\end{figure}
Figure~\ref{fig:profile_ff} shows
the cloud profile averaged over the 4.5~$\mu$s corresponding to the passage of 16 trains of 34 bunches each.
The effect on the cloud profile of a 40-G solenoidal field is shown in Fig.~\ref{fig:profile_ffsol}.

\section{EC Buildup in the Wiggler Magnets}
The EC buildup in the wiggler magnets has been simulated using the CLOUDLAND code~\cite{ref:ipac10lw}. 
The ring length occupied by wigglers is 118~m in the ILC DR lattice design. 
The simulation assumes a peak SEY of 1.2 at an incident electron energy of
250~eV for the 
copper surface of the wiggler vacuum chamber. The absorbed photon rate assumed 
in the simulation is 0.198~photons/m/positron and the azimuthal distribution around the perimeter of the 
vacuum chamber cross section
is approximated as uniform. 
A quantum efficiency of 10\% and rms beam sizes $\sigma_{x}/\sigma_{y} = 80$~$\mu$m/5.5~$\mu$m is
assumed.
The peak wiggler field is 2.1~T. The beam chamber of the wiggler section includes an antechamber 
with 2.0~cm vertical aperture. Our assumption of a round chamber of inner diameter 46~mm is a
reasonable approximation since most electrons accumulate near the vertical midplane due to multipacting. 
The CLOUDLAND calculation shows that a beam with bunch population of $2 \times 10^{10}$ and bunch spacing 
of 6~ns can excite strong multiplication near the vertical midplane. The calculation was performed for
a train of 34~bunches followed by a gap of 45~RF buckets.

\begin{figure}[tb]
   \centering
   \includegraphics*[width=0.99\columnwidth]{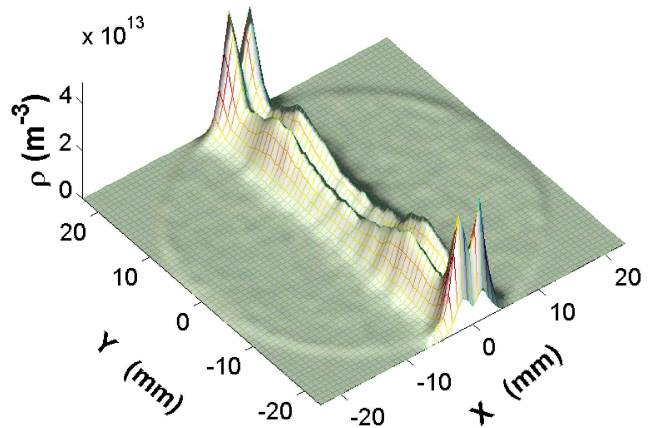}
   \caption{ \label{fig:Wiggler_Profile} 
            Simulated EC distribution in transverse plane at the position of maximum vertical magnetic field
            component in the wiggler for the case of no clearing voltage
}
\end{figure}

Figure~\ref{fig:Wiggler_Profile} shows the distribution of the simulated EC in the transverse plane at the 
longitudinal center of
a wiggler pole for the case of no voltage applied to the clearing electrode. 
This cloud profile is summed over the time corresponding to the passage of 34 bunches followed 
by 45 empty RF buckets. 
The transverse distribution is similar to that found in simulations of cloud buildup in a dipole magnet, 
since photoelectrons produced on the top and bottom of the
vacuum chamber are trapped on the vertical field lines. 
The peak electron density averaged over the beam pipe which is present at arrival of the 
last bunch along the bunch train is about $1.2 \times 10^{13}$~m$^{-3}$. 
The photoelectrons generated at the vertical magnetic field null between poles can contribute a horizontal stripe 
with low density due to the lack of multiplication~\cite{ref:ipac10lw}.
Such a density is negligible compared to that shown here. 
However, such electrons can persist on a time scale long compared to the revolution period due to 
mirroring~\cite{ref:ipac10lw}.

\begin{figure}[tb]
   \centering
   \includegraphics*[width=0.99\columnwidth]{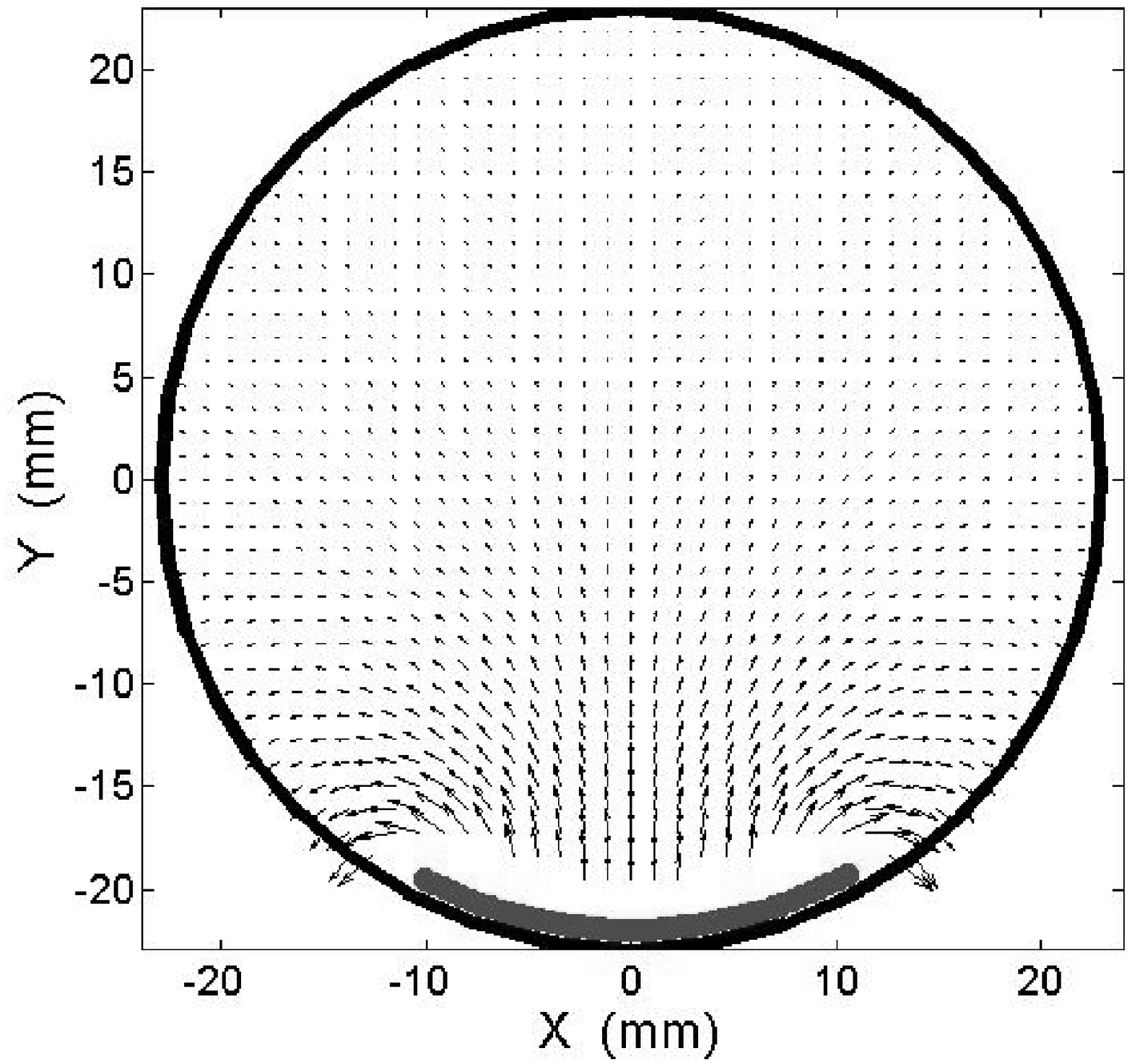}
   \includegraphics*[width=0.99\columnwidth]{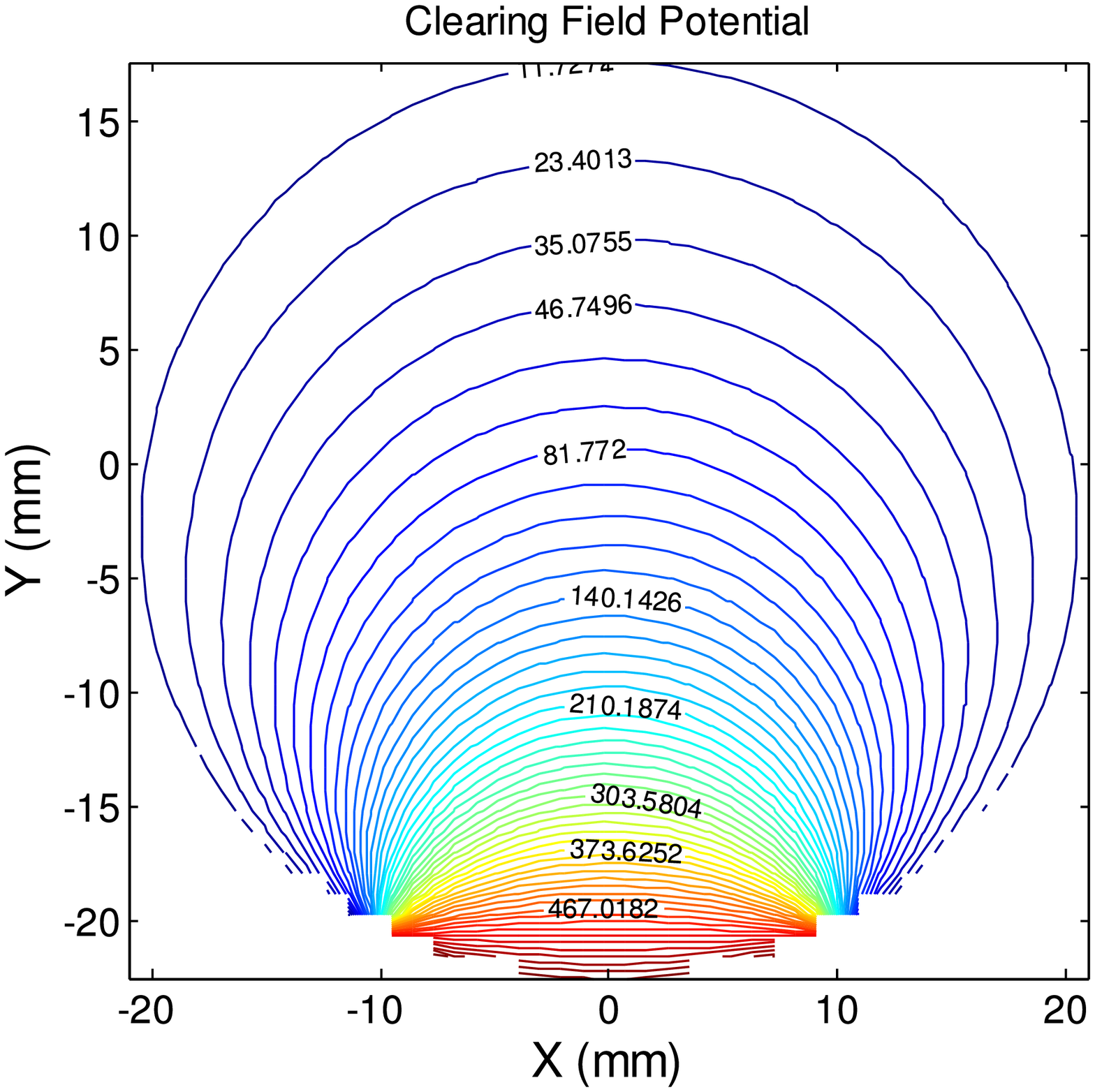}
   \caption{ \label{fig:Clearing_Field} 
            The clearing electrode electric field in the simulated wiggler vacuum chamber
            for an electrode voltage of 500~V.
            The top plot shows the field vectors. The bottom plot shows equipotential lines,
            labeled in units of volts.
}
\end{figure}

The curved clearing electrode foreseen for the wiggler vacuum chambers has
a width of 20~mm and is located on the bottom of the chamber. 
The electrode design consists of a tungsten thermal spray on an 
alumina insulator.  We have conservatively assumed the 
copper SEY parameters for the electrode surface as well.
Figure~\ref{fig:Clearing_Field} shows the field pattern for the simulated clearing electrode.
The potential values on the equipotential lines allow an estimation of the clearing efficiency
for cloud electrons of given kinetic energies.
Since the electrons primarily impact the chamber surface 
near the vertical midplane, the clearing field near that region is 
important for the suppression of electron multiplication. 
Secondary electrons produced between bunch passages
carry energies of just a few electron volts, so
a weak clearing potential is sufficient to prevent them from approaching or leaving the electrode surface.

\begin{figure}[tb]
   \centering
   \includegraphics*[width=0.99\columnwidth]{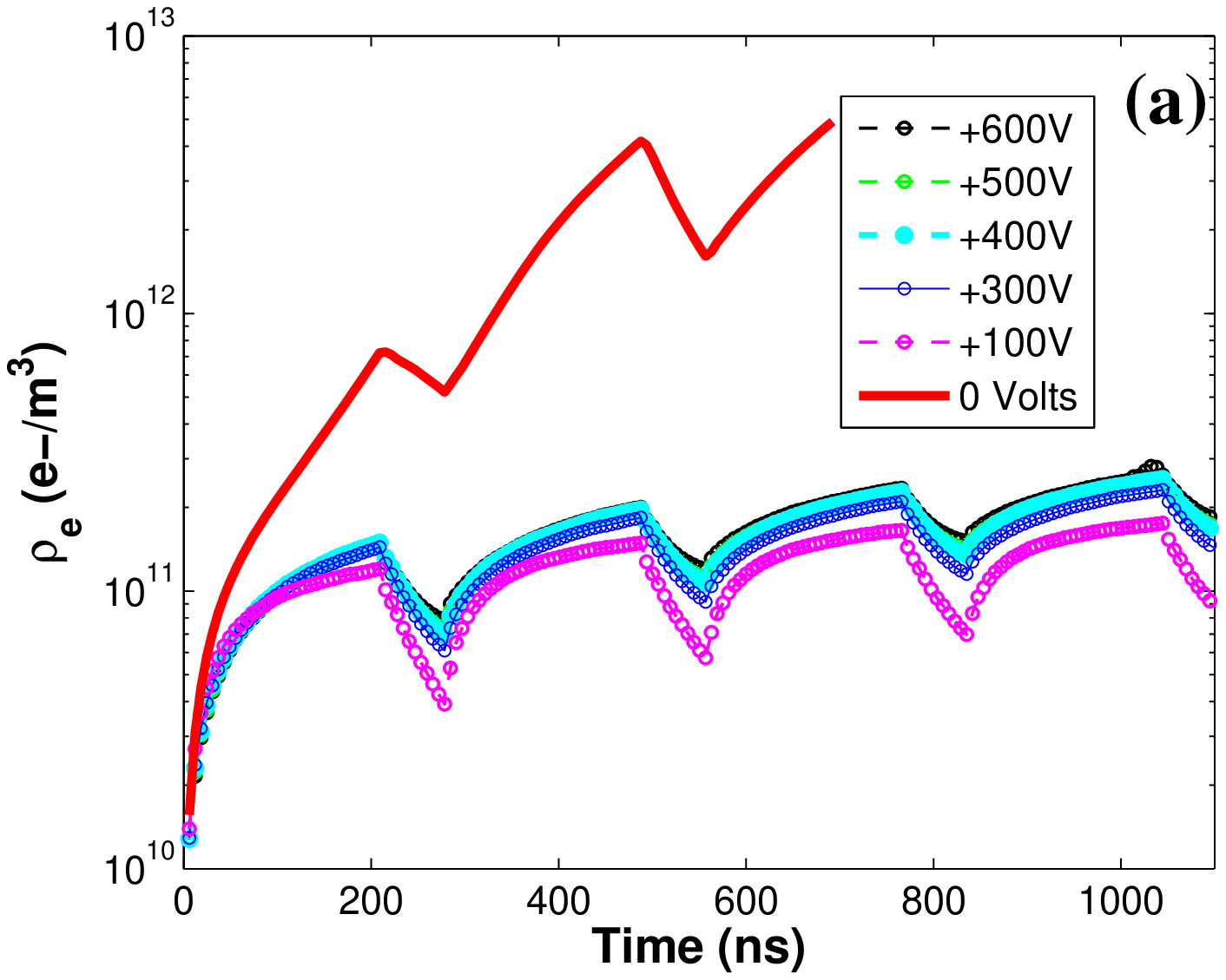}
   \includegraphics*[width=0.99\columnwidth]{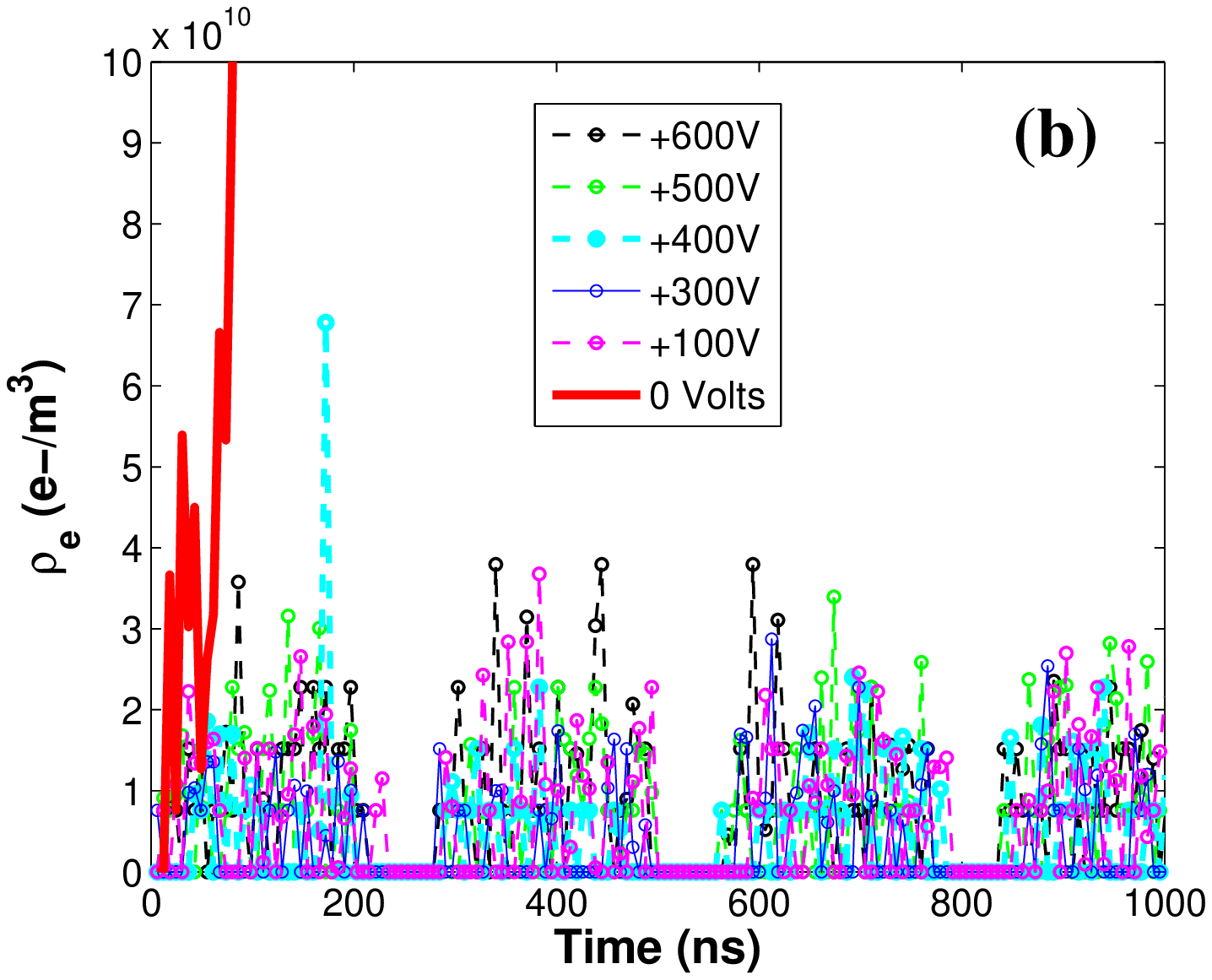}
   \caption{ \label{fig:Wig_Density_Pos} 
            Simulated buildup of the EC at the center of a wiggler pole with a positive 
            clearing voltage applied: 
            a)~cloud density averaged over the vacuum chamber, b)~central cloud density
}
\end{figure}

We simulated electrode voltages from -600~V to +600~V and found that a positive electrode 
bias of 100~V is sufficiently effective at suppressing multipacting. 
Figure~\ref{fig:Wig_Density_Pos} shows the effect of biases up to 600~V. 
With a positive clearing voltage, there are only a small number of macroparticles near the beam, 
so the modeled density shows statistical fluctuations. The density near the positron beam is less than 
$4 \times 10^{10}$~m$^{-3}$, and the density averaged over the vacuum chamber is less than $2 \times 10^{11}$~m$^{-3}$.

\begin{figure}[tb]
   \centering
   \includegraphics*[width=0.99\columnwidth]{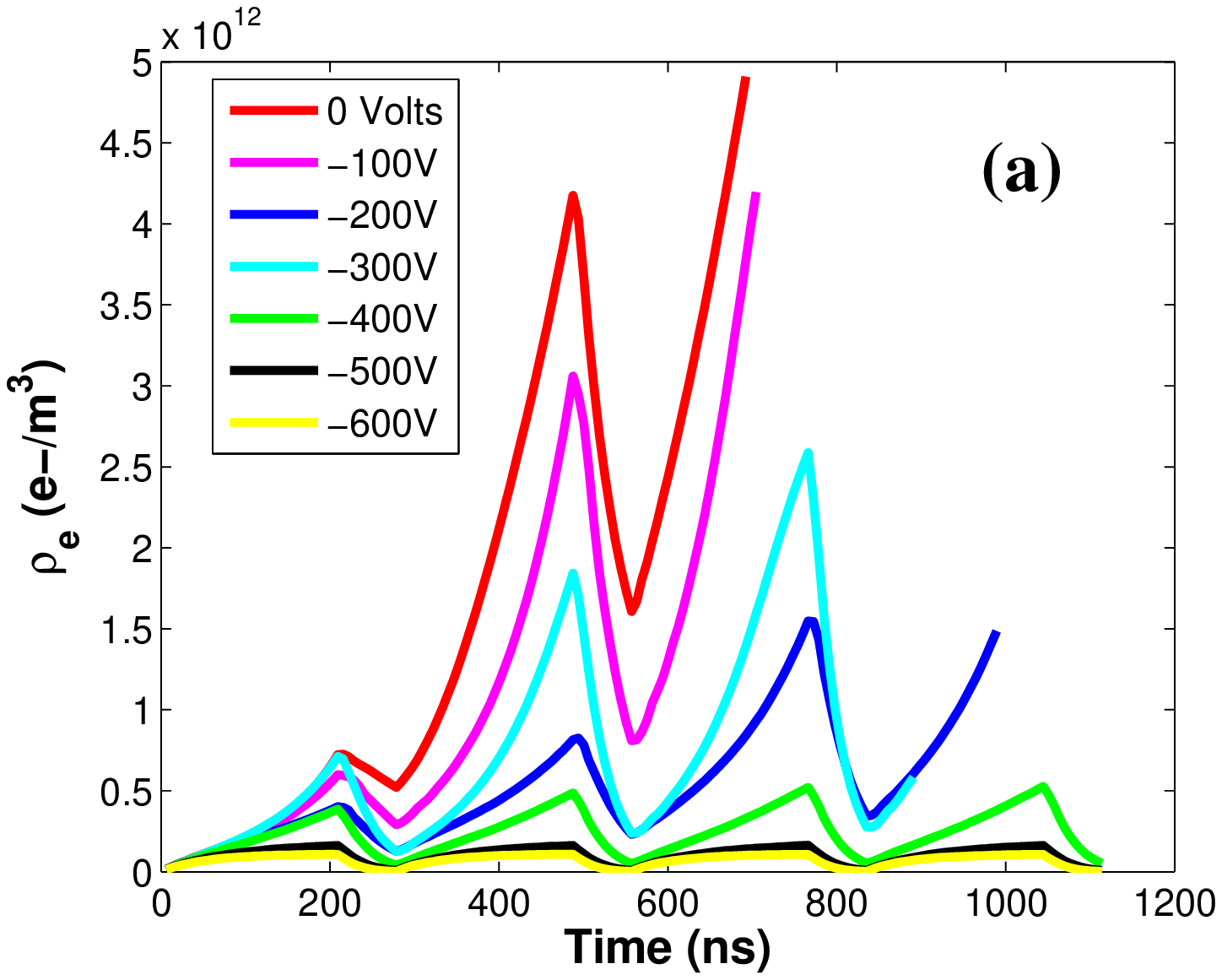}
   \includegraphics*[width=0.99\columnwidth]{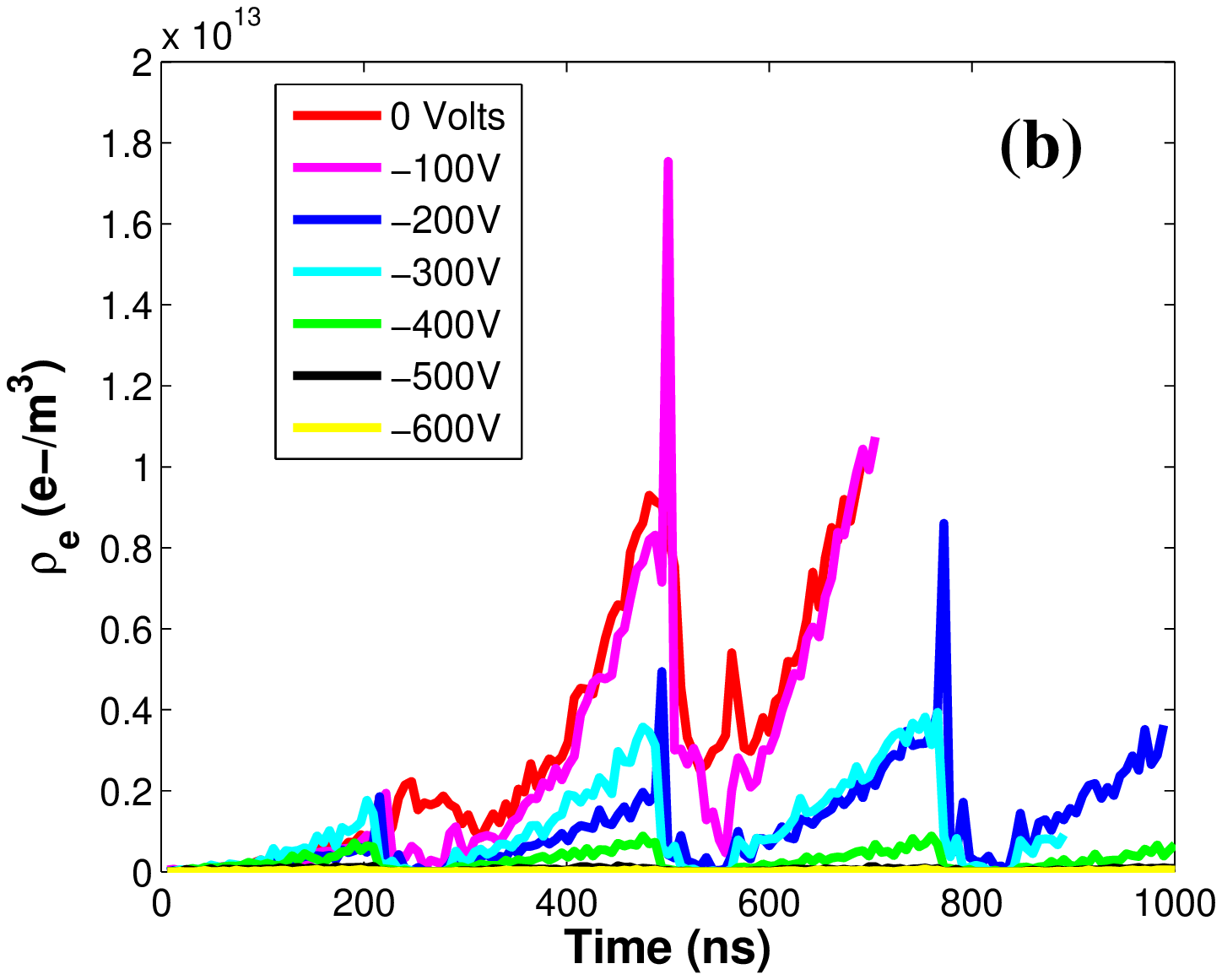}
   \caption{ \label{fig:Wig_Density_Neg}
            Simulated buildup of the EC at the center of a wiggler pole with a negative clearing voltage applied: 
            a)~cloud density averaged over the vacuum chamber, b)~central cloud density
}
\end{figure}

A negative electrode bias also clears cloud electrons, 
but is less effective, as shown in Fig.~\ref{fig:Wig_Density_Neg}, 
especially when the voltage is low. 
A strong field is required to clear the EC.
Interestingly, the suppression is not a monotonic function of the clearing voltage. 
For instance, the average electron density for -300~V is larger than that of -200~V. 
The complicated dynamics due to the clearing field, positron beam kick and space 
charge field accounts for this nonmonotonic dependence.  

\begin{figure*}[tb]
   \centering
   \includegraphics*[width=0.99\columnwidth]{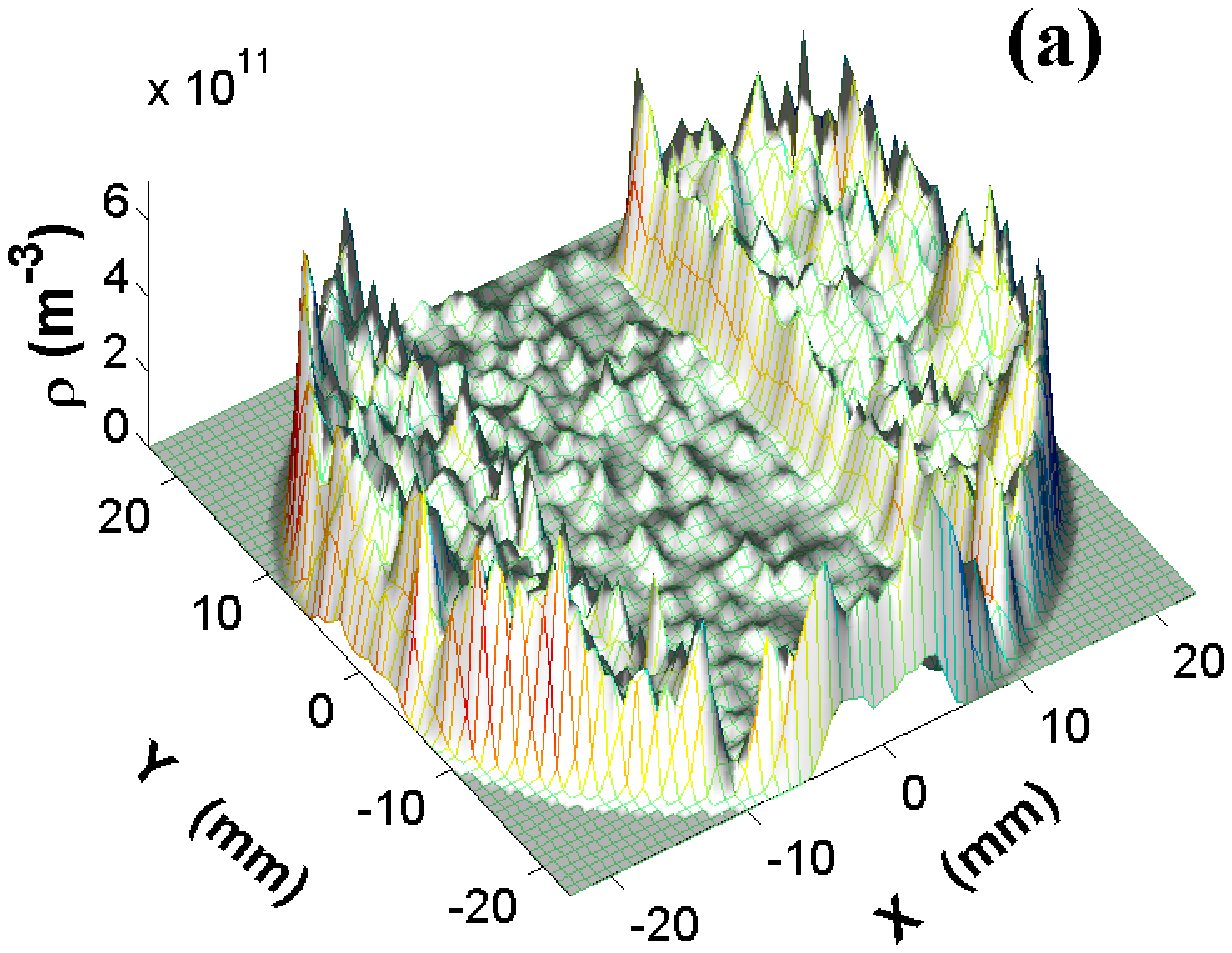}
   \includegraphics*[width=0.99\columnwidth]{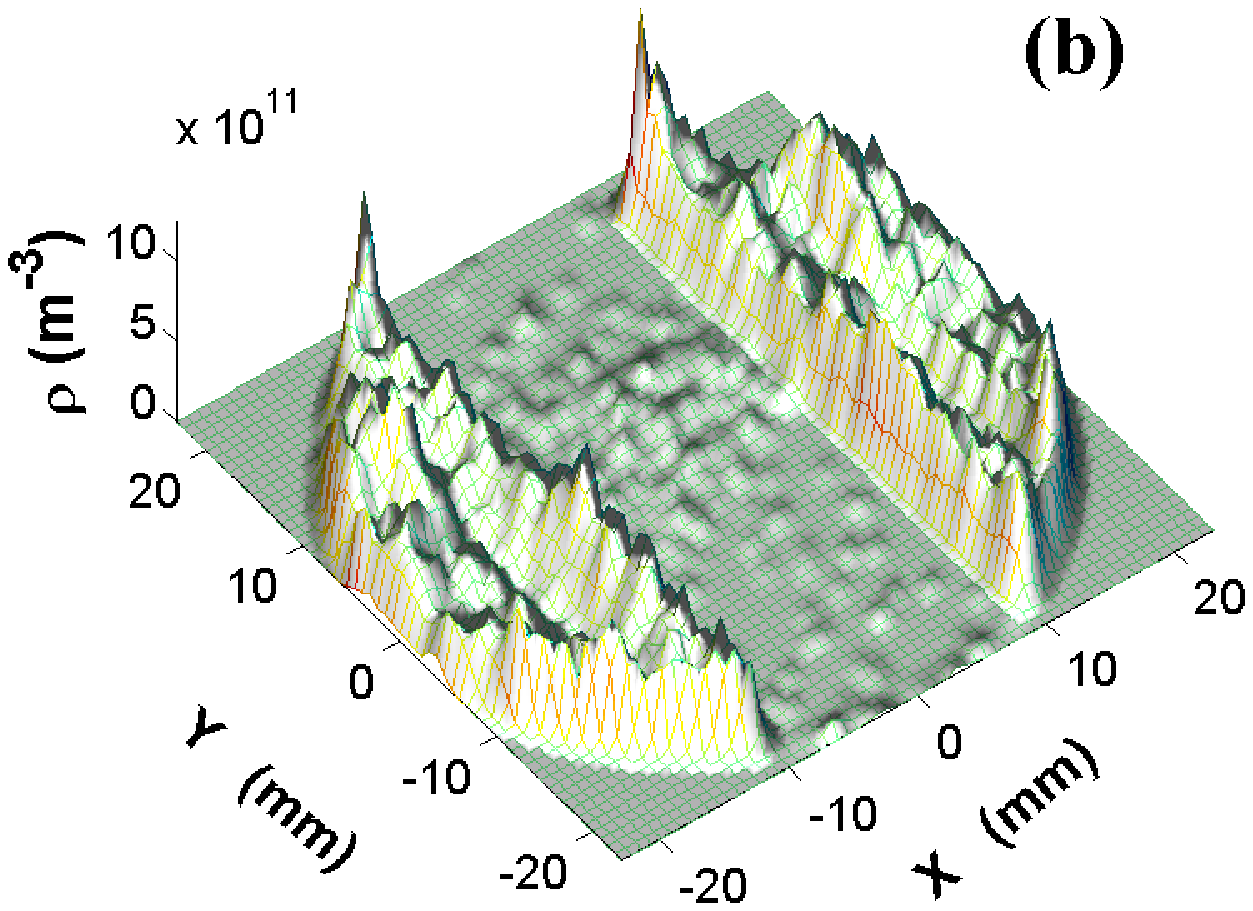}\\
   \includegraphics*[width=0.99\columnwidth]{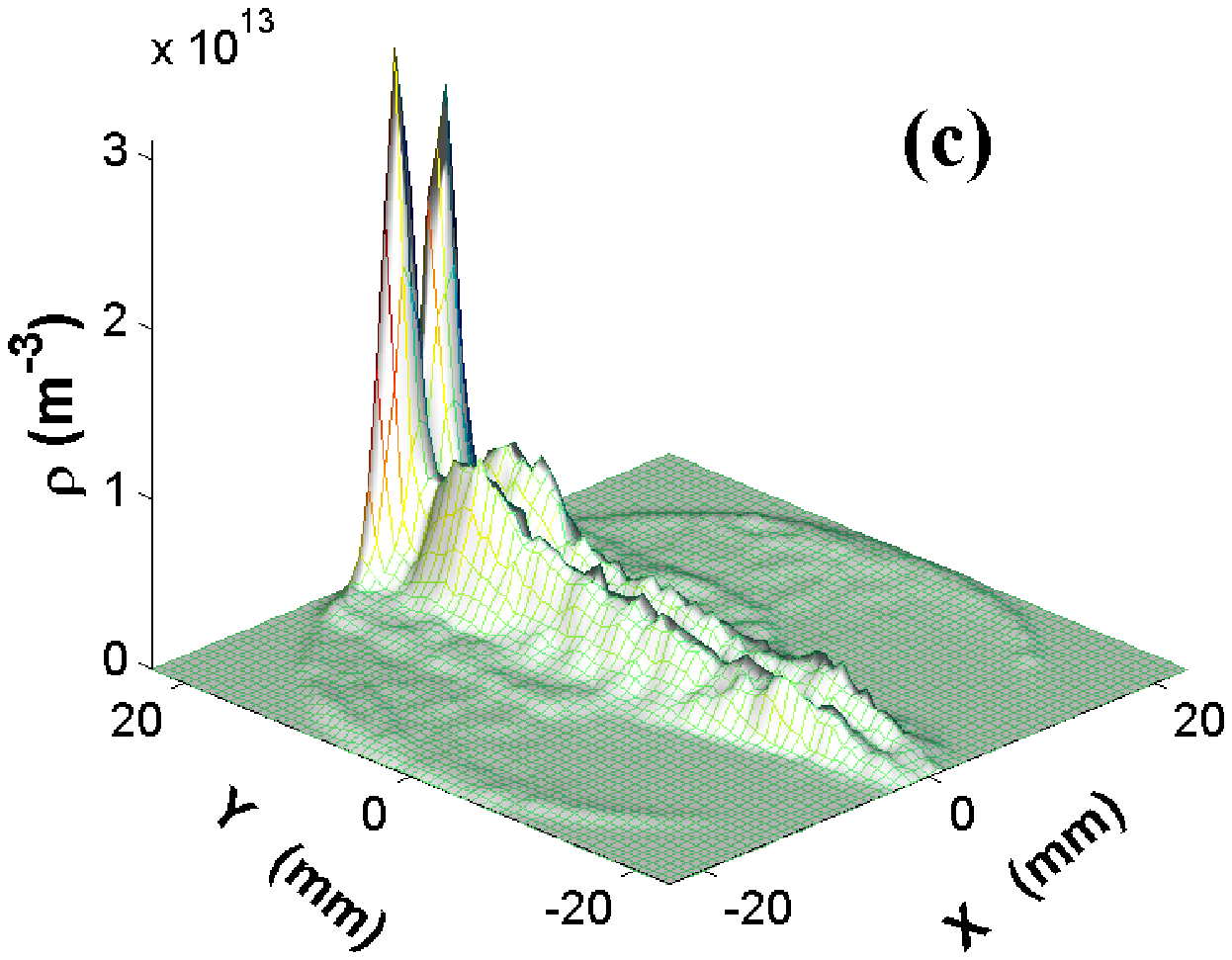}
   \includegraphics*[width=0.99\columnwidth]{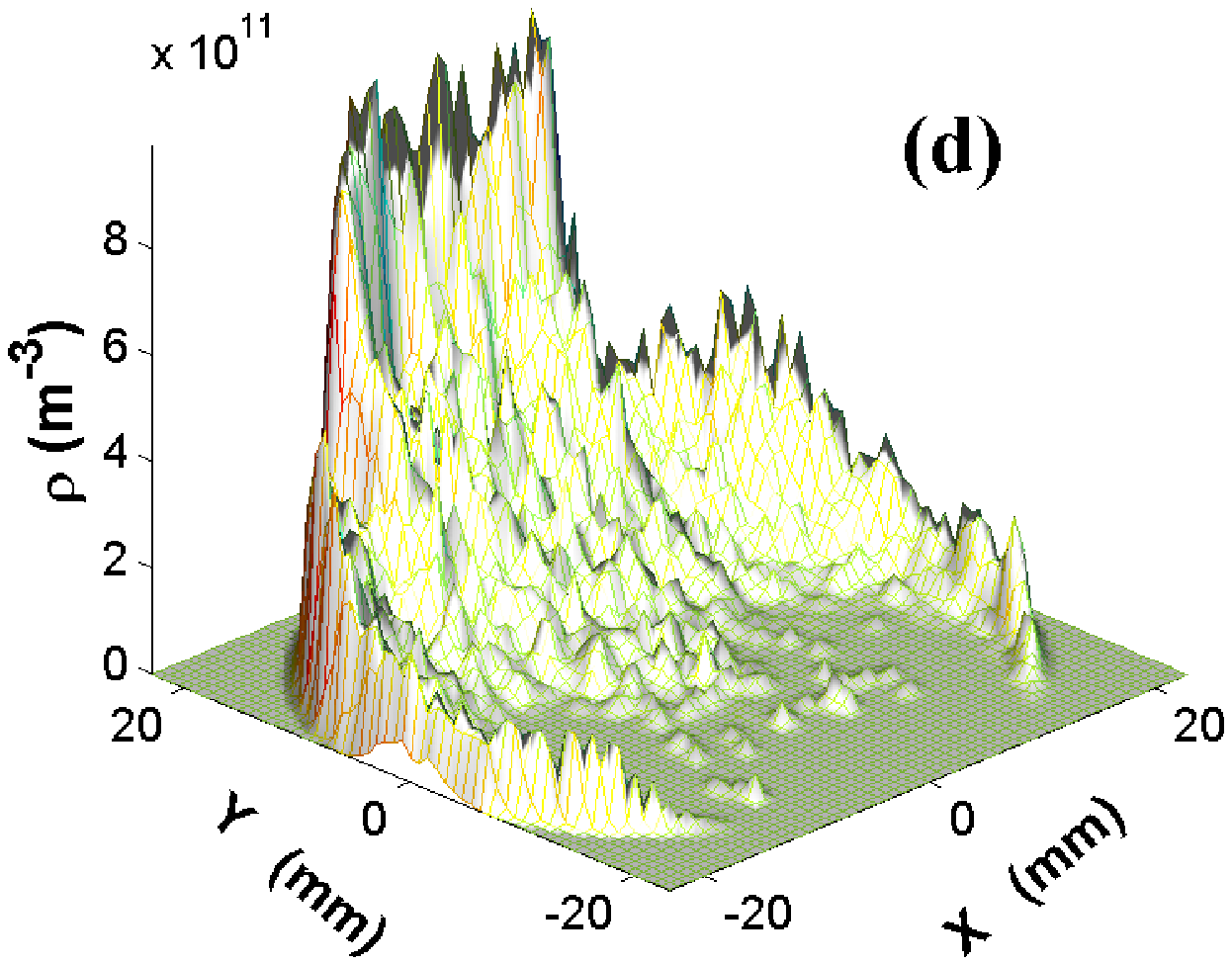}
   \caption{ \label{fig:wigprofiles}
            Simulated EC transverse distributions at a wiggler magnet pole center
            for four values of the clearing voltage:  
            (a)~+100~V; (b)~+600~V; (c)~-300~V; (d)~-600~V.
}
\end{figure*}

The fundamental difference between 
positive and negative clearing voltages is the location where electron multipacting is suppressed. With a 
positive voltage, the photoelectrons and secondary electrons from the electrode surface are 
confined near the surface of the the clearing electrode. 
After a low-energy secondary electron is 
emitted from the electrode, it follows the magnetic field lines upward and is turned back 
to the electrode by the clearing field. The secondary electrons are thus confined 
near the electrode surface. The electrons near the electrode surface can be clearly seen in 
Fig.~\ref{fig:wigprofiles}~a) 
where a weak voltage of 100~V is applied. When the voltage is increased to +600~V, the 
electrons are 
closer to the electrode surface and disappear entirely in Fig.~\ref{fig:wigprofiles}~b). 

Figure~\ref{fig:wigprofiles}~a)  shows that a weak positive voltage results in a low-density
region with the same horizontal width as the electrode. The suppression of the cloud is effective only 
over the horizontal region covered by the electrode. 
The photoelectrons and secondary electrons 
emitted from the top of the chamber spiral downward and reach the
electrode surface. These electrons can have high enough kinetic energies to generate secondary electrons 
due to the acceleration imparted by the positron beam. However, secondary electrons thus produced will generally 
be trapped by the clearing field. Therefore, a positive bias effectively collects
both the photoelectrons and secondary electrons.

In the case of a negative clearing voltage, both photoelectrons and secondary electrons can leave the electrode surface. 
The electrons emitted from the electrode surface are accelerated toward the beam axis and 
end up collecting near the top of the chamber, as can be 
seen in Figs.~\ref{fig:wigprofiles}~c) and~d). 
Although this results in clearing the cloud from the beam region, the clearing 
field near the top of the surface is much weaker than near the electrode and the field lines deviate from the 
vertical direction as shown in Fig.~\ref{fig:Clearing_Field}. 
This makes negative biases much less effective
for clearing the cloud, so a much stronger field is required. 
A voltage of -300~V does not suppress the 
multiplication. Even a voltage of -600~V does not significantly reduce the density of electrons near the 
center of beam pipe and there remain a large number of electrons near the upper surface.

\section{Beam Dynamics Simulation}
Single-bunch instabilities and the 
dilution of vertical emittance is a primary concern of EC effects in 
DRs~\cite{ref:IPAC12:WEYA02,ref:JJAP50:026401,PRSTAB7:124801}. 
The modeling work on EC buildup described above provides estimates of the cloud density
in the region near the beam at the arrival times of the bunches. 
The estimates place an upper limit
on the ring-averaged density of about $3.5 \times 10^{10}$~m$^{-3}$.
The additional cloud buildup suppression provided by the grooved surfaces recommended for the arc 
dipole regions has yet to be calculated for the ILC DR lattice.
Based on these results for upper limits on the cloud densities, the
simulation code CMAD~\cite{ref:cmad} has been used to estimate single-bunch
instability thresholds and emittance dilution arising from the beam-cloud interaction.
The parameters used in these simulations 
were based on the DTC03 lattice design. The study was performed with two models 
of the ring beta functions. The first used a continuous-focusing model. 
The second involved the full lattice of the DR. The continuous-focusing model is highly simplified 
but is far more efficient in performing computations. This model was used to scan through a set 
of cloud densities in order to  estimate the range over which the behavior transitions from  
gradual and linear to a fast exponential growth in emittance. It should be noted that effects of 
damping and diffusion due to synchrotron radiation emission are not included in the
calculations. The time scales of radiation damping and quantum excitation are both much less than 
the instability growth time.

The CMAD simulation algorithms are similar to those of other programs such as HEADTAIL~\cite{ref:headtail}, 
WARP~\cite{ref:warpQS} and PEHTS~\cite{ref:phets}. Results from CMAD, HEADTAIL and WARP have been compared 
for the the continuous-focusing and full lattice cases~\cite{ref:code_comp1, ref:code_comp2}. 
The continuous-focusing model uses a constant beta function value that is obtained from the betatron tunes and the circumference. 
The model has no dispersion, so there is no variation of the beam size around the simulated ring.
The model uses a number of 
beam-cloud interaction points (IPs) around the ring
sufficient to avoid artificial resonances arising from the discreteness of 
beam-cloud interaction.   

The details of the physical and computational parameters are given in Tabs.~~\ref{tab:cmad_physparams} 
and~\ref{tab:cmad_compparams} respectively.
\begin{table}
\caption{\label{tab:cmad_physparams}
         List of physical parameters used in the CMAD simulations, corresponding 
         to the ILC DR lattice design}
\begin{tabular}{lc}
\toprule[1pt]
\addlinespace[2pt]
 Beam energy & 5 GeV   \\ 
 Unnormalized emittance x, y &  0.5676~nm, 2.0 pm \\ 
 Bunch population & 2 $\times 10^{10}$  \\ 
 Bunch length & 0.6036 cm \\  
 Tunes x, y, z &  48.248, 26.63, 0.0314 \\ 
 Momentum compaction & 3.301$\times 10^{-3}$  \\ 
 Circumference & 3234.3540 m  \\ 
 Energy spread & $1.1 \times 10^{-3}$ \\ 
 Chromaticity ($\xi_{\rm x}=\xi_{\rm y}$) & 1.0  \\ 
\bottomrule[1pt]
\end{tabular}
\end{table}
\begin{table}
\caption{\label{tab:cmad_compparams}
         List of computational parameters used in the CMAD simulations. The number of IPs is
         used only in the continuous-focusing model.}
\begin{tabular}{lc}
\toprule[1pt]
\addlinespace[2pt]
Macro e+ &300000 \\  
Macro e- &100000 \\ 
Bunch slices& 96 \\ 
Grid nodes &128 $\times$ 128 \\ 
Domain extent x, y & 20 sigma \\ 
Domain extent z & +/- 2 sigma \\ 
IPs (uniform $\beta$ only)) & 400 \\ 
Nr processors used in parallel & 96 \\ 
\bottomrule[1pt]
\end{tabular}
\end{table}
Our simulations assume chromaticity values typical of 
storage rings such as CESR.
Collective effects require that the chromaticity
be set to a reasonable value to ensure stability via chromatic damping.  
The computational parameters 
were chosen based on experience with simulations
for {\cesrta} \cite{ref:sonnad_cmad1,ref:sonnad_cmad2}. 
The computational domain was truncated at 20 rms beam sizes in the transverse directions and at 2 rms beam sizes in either 
direction for the longitudinal extent. The large transverse extent ensures that sufficient EC is included
to accurately model the pinching process.
The number of IPs used in the continuous-focusing
model was 400, much greater than the number of betatron oscillations per revolution (see Tab.~\ref{tab:cmad_physparams}). 
The beam was sliced longitudinally into 96 segments and the computation 
was performed in parallel, distributed over 96 processors. 

\begin{figure}[tb]
    \centering
    \includegraphics*[width=0.9\columnwidth]{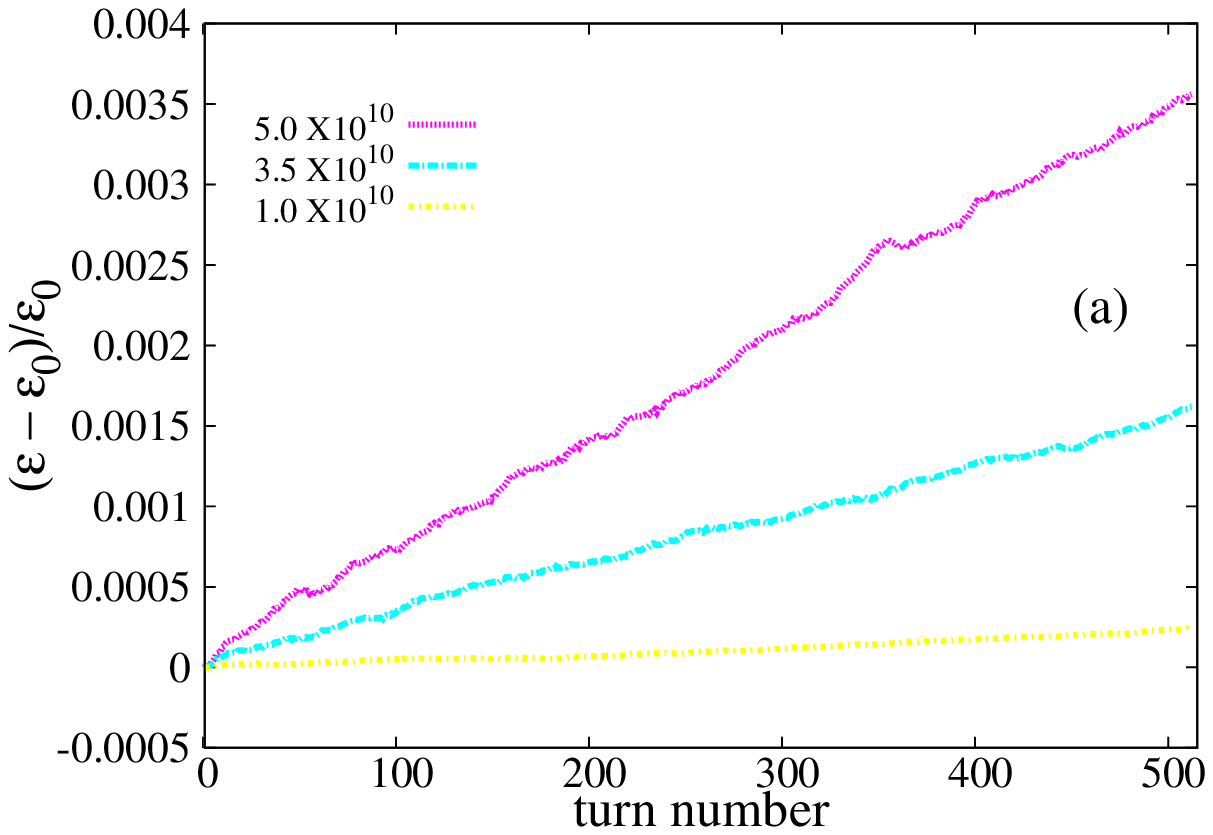}
    \includegraphics*[width=0.9\columnwidth]{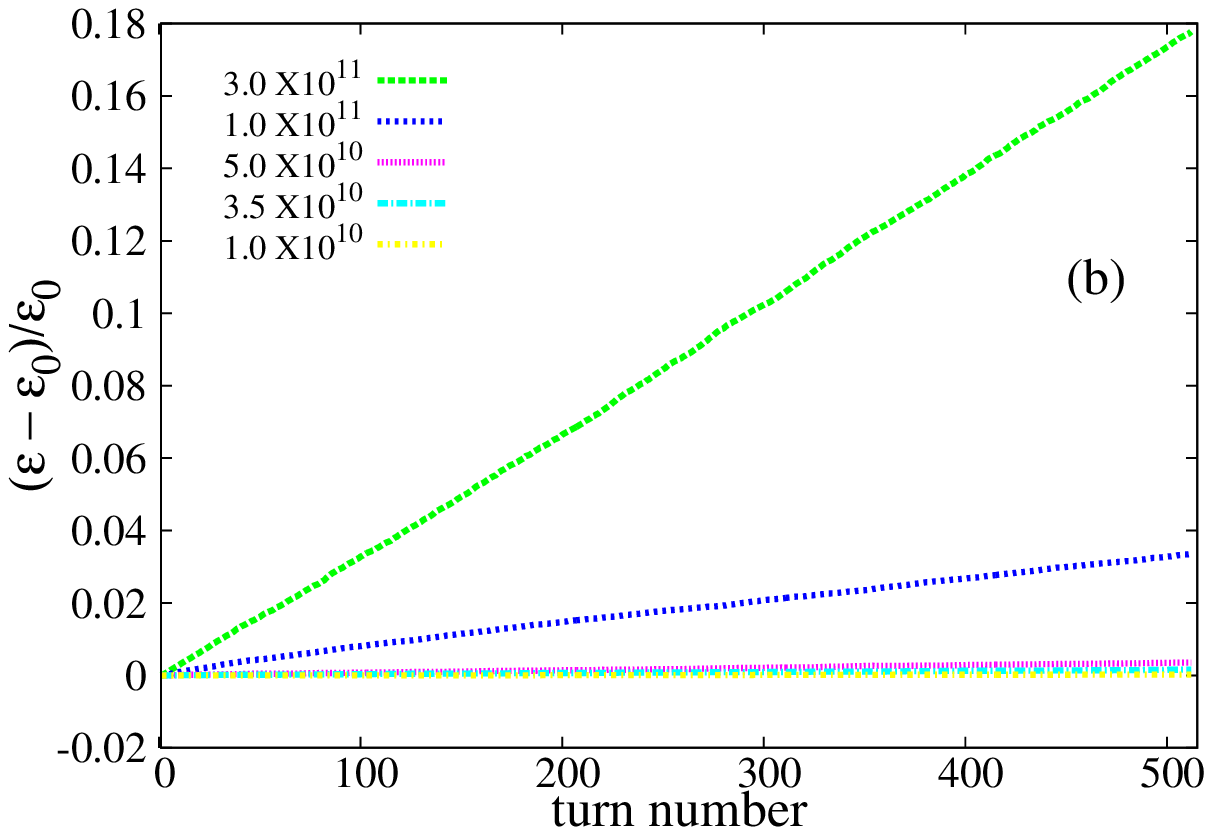}
    \includegraphics*[width=0.9\columnwidth]{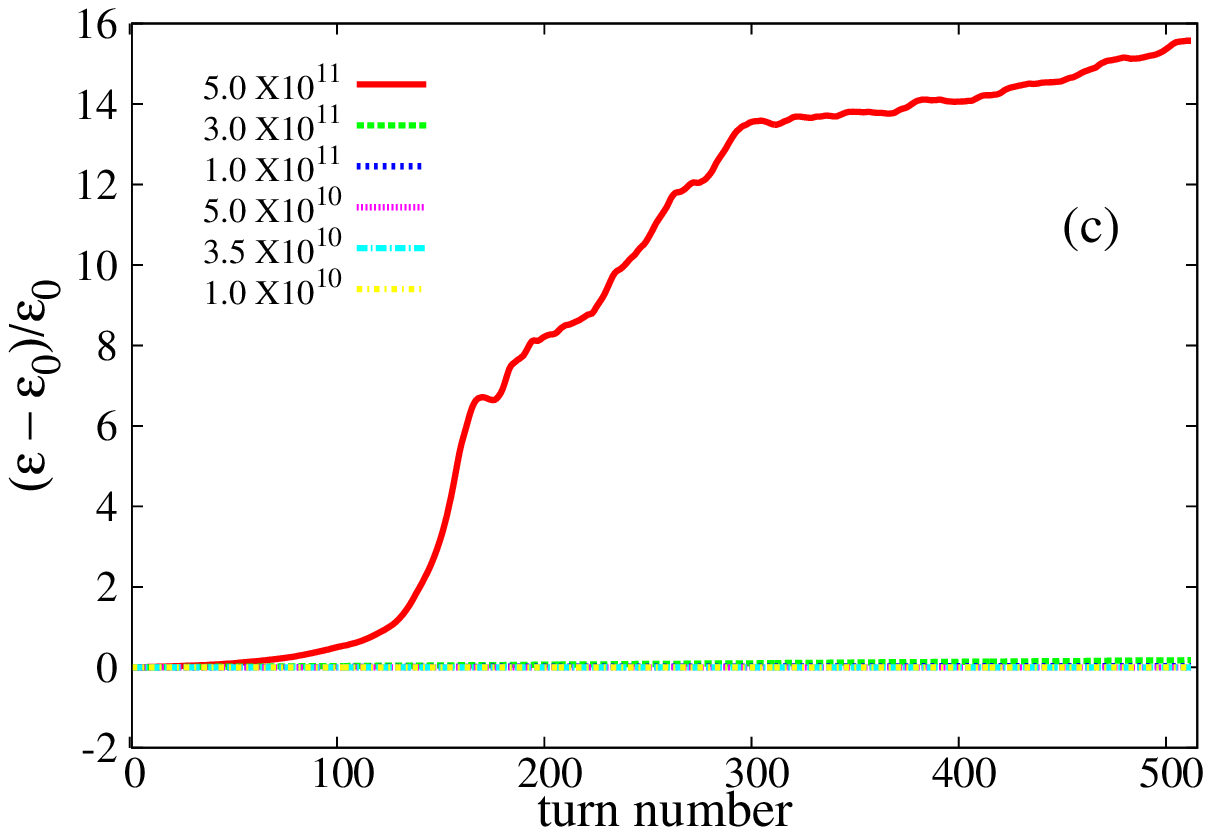}
    \caption{\label{fig:cmad_smooth_focus}
             Modeled relative vertical emittance growth using the continuous-focusing model for
             various cloud densities. a)~The relative growth rate over 500 turns is 0.16\% for 
             an average density of $3.5 \times 10^{10}$~m$^{-3}$, which is the value
             determined by the cloud buildup modeling for the ILC DR. 
             b)~The growth rate increases to 18\% in 500 turns for a density of 
             $3.0 \times 10^{11}$~m$^{-3}$.
             c)~Exponential growth is found for a density of $5.0 \times 10^{11}$~m$^{-3}$.
} 
\end{figure}

Figure~\ref{fig:cmad_smooth_focus} shows the calculated
emittance growth over a period of 500~turns using the continuous-focusing model. 
The simulations were done for EC densities ranging 
from $1.0 \times 10^{10}$~m$^{-3}$ to $5.0 \times 10^{11}$~m$^{-3}$. Figure~\ref{fig:cmad_smooth_focus}~a) 
shows the 
emittance growth rate for three cases, with the intermediate cloud density of $3.5 \times 10^{10}$~m$^{-3}$ 
corresponding to the estimated ring-averaged cloud density 
in the vicinity of the beam. 
Increasing the cloud density to $5.0 \times 10^{10}$~m$^{-3}$ results in
no deviation from a linear dependence on turn number. 
Figure~\ref{fig:cmad_smooth_focus}~b) shows that the vertical emittance 
growth rate increases almost by two orders of magnitude when the cloud density increases 
from  $3.5 \times 10^{10}$~m$^{-3}$ to $3.0 \times 10^{11}$~m$^{-3}$. 
Figure~\ref{fig:cmad_smooth_focus}~c) shows that the growth rate transitions from 
linear to exponential when the density is raised to $5.0 \times 10^{11}$~m$^{-3}$. 
The linear region below the ``instability threshold'' has been observed 
in  single-bunch simulations with other modeling codes~\cite{ref:PRL97:034801,ref:JJAP50:026401}. 
The operating conditions of the ring must be kept well below the transition from linear to exponential
dependence in order to ensure stable operation.
Our results show that the cloud density for the ILC DR operating conditions can be expected 
to be an order of magnitude below this transition point.
\begin{table}
\caption{\label{tab:cmad_cloudparams}
         Cloud density in the ring elements and their occupancy fractions}
\begin{tabular}{lcc}
\toprule[1pt]
\addlinespace[2pt]
Element & Cloud density & Occupancy (\%) \\ 
Fieldfree & 0  & 66 \\ 
Dipoles & $4.0 \times 10^{10}$ & 15.14 \\ 
Quads in arcs & $1.6 \times 10^{11}$ & 9.8 \\ 
Sextupoles in arcs & $1.4 \times 10^{11}$ & 5.56 \\ 
Wigglers & $1.5 \times 10^{10}$ & 2.96 \\ 
Quads in wiggler region & $1.2 \times 10^{12}$ & 0.49 \\ 
Average & $3.5 \times 10^{10}$ & \\ 
\bottomrule[1pt]
\end{tabular}
\end{table}

Estimates of emittance growth were also calculated using the full lattice 
design of the DR. 
The beam particles were transported using first order $6 \times 6$ transfer matrices, thus including
variation of the horizontal and vertical beam size with beta function and dispersion. 
In particular, the beam size ratio ${\sigma}_x/{\sigma}_y$ reached a value of about 100, imposing challenging 
numerical accuracy conditions. 
This was overcome by altering the Poisson solver at points with a beam aspect ratio higher than~20. 
The beam underwent an
interaction with the EC at each element in the lattice. Thus the number of IPs used in this case
was 5765, equal to the number of elements in the lattice design model. The wigglers were modeled using a 
bend-drift-bend
sequence. 
Electrons in regions
with an applied magnetic field, including those in the wiggler sections, were tracked based on the full 
Lorentz force exerted
on the particle. The influence of the external field influences the pinching process, a feature that 
is missing in the
continuous-focusing calculations. 
The cloud density in each element was set to
the value derived from the buildup simulations.
These densities 
are listed in Tab.~\ref{tab:cmad_cloudparams}. Thus, several physical 
details omitted from the continuous-focusing model were 
taken into account in this simulation of the full lattice. 

\begin{figure}[htb]
\includegraphics*[width=0.99\columnwidth]{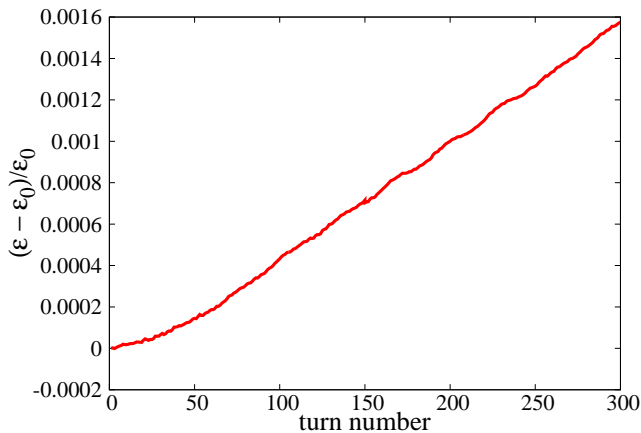}
\caption{Estimate of the emittance growth using the full lattice in the model with
specified cloud densities in each element of the lattice}
\label{fig:cmad_lattice}
\end{figure}

Figure~\ref{fig:cmad_lattice} shows the evolution of the beam emittance under the EC conditions given 
in Tab.~\ref{tab:cmad_cloudparams}.
The calculation estimates the relative emittance growth over 300 turns to be 0.16\%. 
Except for an initial transient 
phase in the first 25 turns, the growth is linear. The ILC DR beam store time is 18550 
turns. In the absence of any damping mechanism over this time period, one can therefore expect the
beam emittance to increase by 10\% due to ECs during the store time.
When the same extrapolation is applied to the continuous-focusing case with a
cloud density of $3.5 \times 10^{10}$~m$^{-3}$, we obtain a growth of 6\% in beam emittance 
during 18550 turns.  

%This conclusion that such cloud densities are well below the head-tail instability
%threshold is further supported by the {\cesrta} measurements of head-tail excitation lines 
%at 2.1 and 4.0~GeV, 
%from which one can conclude that the threshold cloud density at 5~GeV must 
%exceed \mbox{$2.0 \times 10^{12}$~m$^{-3}$}~\cite{ref:IPAC12:WEYA02}.

\begin{figure}[htb]
\includegraphics*[width=0.99\columnwidth]{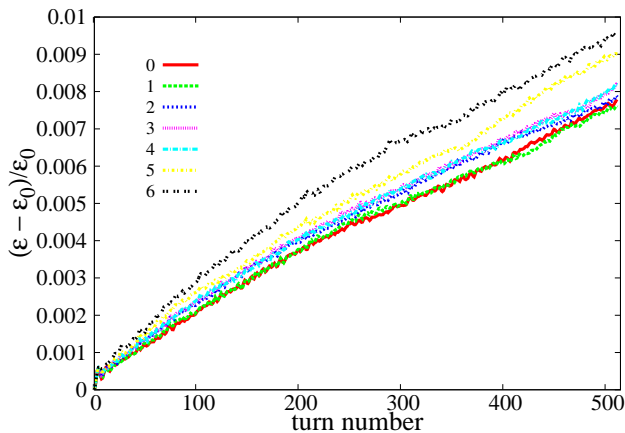}
\caption{Calculation of emittance growth for  chromaticity values ranging
from 0 to 6 in units of $\xi=dQ/d{\delta}$ where $\delta$ is the relative energy spread
in the positron beam. These calculations were performed using the continuous-focusing
model with a ring-averaged cloud density of $3.5 \times 10^{10}$~m$^{-3}$. }
\label{fig:chrom_scan}
\end{figure}

We have investigated the dependence of this result on the chosen chromaticity.
%The chromaticity in the lattice was not part of the original design. The value 
%chosen was typical of what some other accelrators such as the CesrTA operate at. 
%The avoidance of collective effects requires that the chromaticity be set to a reasonable 
%value to ensure stability via chromatic damping. 
%In order to evaluate 
%the effect of increased chromaticity, computations were performed using the continuous-focusing 
Computations were performed using the continuous-focusing 
model with the cloud density of 
$3.5 \times 10^{10}$~m$^{-3}$ given by the buildup simulations. 
For the purposes of this investigation of the chromaticity dependence, 40 beam-cloud interaction
points were used, rather than the 400 modeled in the full simulation.
Figure~\ref{fig:chrom_scan} shows that the chromaticity
influences the emittance growth only moderately for this cloud density.
The calculated emittance grows from 0.8\% to 1.0\% as the chromaticity increases from~0 to~6.

Future work on the beam dynamics simulation will include modeling of the detailed
vacuum chamber cross section.
Rather than using a uniform distribution of cloud electrons,
the initial electron cloud distribution for these simulations should be imported from 
the buildup calculations. In addition, the tracking of the beam should be performed 
for the full store period of the damping ring and include the damping and diffusion effects.

\section{Summary}
We have updated the lattice design for the 3.2-km, 5~GeV ILC positron damping ring and calculated 
the 
distributions of synchrotron radiation around the ring, including the effects of photon scattering 
inside the 
vacuum chamber. This analysis was used to refine the choice of electron-cloud-mitigating techniques 
in the various 
magnetic field environments of the arcs and straights. Groove patterns and antechambers were used as 
mitigation techniques
in the modeled dipole magnets, along with TiN-coating in the 5-cm-diameter quadrupole and sextupole
 magnet 
vacuum chambers.
The drift regions were assumed to be equipped with solenoid windings and the wigglers with clearing electrodes.
Electron buildup modeling codes tuned to the measurements of cloud buildup performed at the CESR Test 
Accelerator
were employed to make 
quantitative estimates of the cloud densities near the beam axis at the arrival time of each of the 
6-mm-long 
bunches for the operational bunch configuration of 34-bunch trains separated by 20~m, the bunches 
spaced 1.8~m apart, 
each carrying $2 \times 10^{10}$ positrons. 
An upper limit on the ring-averaged electron density was found of
$3.5 \times 10^{10}$~m$^{-3}$. The cloud densities in the various ring sections then served as 
input to simulations 
of their effect on the positron beam emittance. The calculated emittance growth for these operating 
conditions was 
found to grow linearly with turn number, showing that operation was well below the instability 
threshold. 
Total vertical emittance growth during the entire store time of 18550 turns was found to be about 10\%. 
We can therefore positively assess the operational feasibility of the ILC positron damping ring as 
specified in the 
Technical Design Report. This work will serve as a baseline for future 
optics development, vacuum chamber designs and operating parameters of the ring.

\section{Acknowledgments}
The authors wish to acknowledge important contributions from the technical staffs of the
collaborating laboratories. Our work also benefited from useful discussions with S.~Calatroni, 
J.~Calvey, R.~Cimino, P.~Costa Pinto, 
T.~Demma, K.~Kanazawa, S.~Kato, Y.~Li, K.~Ohmi, G.~Rumolo, Y.~Suetsugu, M.~Taborelli, M.~Venturini,
C.~Yin Vallgren and F.~Zimmermann.
This work is supported by National Science Foundation and
by the US Department of Energy under contract numbers
\mbox{PHY-0734867}, \mbox{PHY-1002467} and \mbox{DE-FC02-08ER41538} and \mbox{DE-SC0006505}.
The beam dynamics simulations also used resources of the National Energy Research Scientific Computing Center,
which is supported by the Office of Science of the U.S. Department of Energy under
Contract No. DE-AC02-05CH11231.

\nocite{*}

\bibliographystyle{apsrev4-1}
\bibliography{ilcdr_ecloud_sim}
\end{document}